\documentclass[11pt]{article}
\usepackage{amssymb}
\usepackage{epsf}

\newtheorem{theorem}{Theorem}[section]

\newtheorem{definition}[theorem]{Definition}

\newtheorem{claim}[theorem]{Claim}
\newtheorem{lemma}[theorem]{Lemma}

\newtheorem{corollary}[theorem]{Corollary}

\newcommand{\qedsymb}{\hfill{\rule{2mm}{2mm}}}
\newenvironment{proof}[1][]{\begin{trivlist}
\item[\hspace{\labelsep}{\bf\noindent Proof#1:\/}] }{\qedsymb\end{trivlist}}

\def\calA{{\cal A}}
\def\calB{{\cal B}}
\def\calC{{\cal C}}
\def\calP{{\cal P}}
\def\calT{{\cal T}}
\def\Z{{\mathbb{Z}}}
\def\R{\mathbb{R}}
\def\N{\mathbb{N}}
\def\mod{\mathrm{mod}}
\def\vol{\mathrm{vol}}
\def\frc{\mathrm{frc}}
\def\diam{\mathrm{diam}}
\def\cc{{\tilde{c}}}
\def\th{{\tilde{h}}}

\newcommand\floor[1]{{\lfloor #1 \rfloor}}
\newcommand\ceil[1]{{\lceil #1 \rceil}}
\newcommand\round[1]{{\lfloor #1 \rceil}}

\newcommand\abs[1]{{\left| {#1} \right|}}
\newcommand\ip[1]{{\langle {#1} \rangle}}
\newcommand\norm[1]{{\| #1 \|}}

\newcommand\C[1]{{ c_{\sf {#1}} }}

\newcommand{\onote}[1]{}

\begin{document}

\title{\bf New Lattice Based Cryptographic Constructions}

\author{
Oded Regev
\thanks{
   EECS Department, UC Berkeley, Berkeley, CA 94720. Email: odedr@cs.berkeley.edu.
   Most of this work was done while the author was at the
   Institute for Advanced Study, Princeton, NJ.
   Work supported by the Army Research Office grant DAAD19-03-1-0082 and
   NSF grant CCR-9987845.}
   }


\maketitle

\begin{abstract}
We introduce the use of Fourier analysis on lattices as an integral part of a lattice based construction. The tools we
develop provide an elegant description of certain Gaussian distributions around lattice points. Our results include two
cryptographic constructions which are based on the worst-case hardness of the unique shortest vector problem. The main
result is a new public key cryptosystem whose security guarantee is considerably stronger than previous results
($O(n^{1.5})$ instead of $O(n^7)$). This provides the first alternative to Ajtai and Dwork's original 1996
cryptosystem. Our second result is a family of collision resistant hash functions which, apart from improving the
security in terms of the unique shortest vector problem, is also the first example of an analysis which is not based on
Ajtai's iterative step. Surprisingly, both results are derived from one theorem which presents two indistinguishable
distributions on the segment $[0,1)$. It seems that this theorem can have further applications and as an example we
mention how it can be used to solve an open problem related to quantum computation.
\end{abstract}

\section{Introduction}

Cryptographic constructions based on lattices have attracted considerable interest in recent years. The main reason is
that, unlike many other cryptographic constructions, lattice based constructions can be based on the {\em worst-case}
hardness of a problem. That is, breaking them would imply a solution to {\em any} instance of a certain lattice
problem. In this paper we will be interested in the unique shortest vector problem (uSVP), a lattice problem which is
believed to be hard: we are asked to find the shortest vector in an $n$-dimensional lattice with the promise that it is
shorter by a factor of $n^c$ than all other non-parallel vectors. Hence, the problem becomes harder as $c$ decreases.
The results in this field can be divided into two types. The first includes public key cryptosystems and the second
includes families of collision resistant hash functions.

The only previously known public key cryptosystem based on a worst-case lattice problem is the one due to Ajtai and
Dwork~\cite{AjtaiDwork} which appeared in 1996. They presented a public key cryptosystem based on the worst-case
hardness of $O(n^8)$-uSVP. Then, in \cite{GoldreichGH97}, Goldreich, Goldwasser and Halevi showed how to eliminate
decryption errors that existed in the original scheme. They also improved the security to $O(n^7)$-uSVP. Although there
are other lattice based cryptosystems (see, e.g., \cite{GoldreichGH97b,NTRU,Micciancio01hnf}), none of them is based on
the worst-case hardness of a lattice problem. Our main result is a new public key cryptosystem whose security is based
on $O(n^{1.5})$-uSVP.

In \cite{Ajtai96}, Ajtai presented a family of one-way hash functions based on the worst-case hardness of several
lattice problems. In terms of the uSVP, it was based on the hardness of $O(n^{c})$-uSVP. The constant $c$ was not
explicitly specified but later it was noted to be $c=19$~\cite{CaiAW}. In \cite{GolGolHal96}, it was shown that under
the same assumptions one can obtain a family of collision resistant hash functions. This is a stronger primitive than a
one-way function with many uses in cryptography. Cai and Nerurkar \cite{CaiNerurkarAW} improved the exponent to
$c=9+\epsilon$ and later, by providing an improved analysis, Cai \cite{CaiAW} obtained $c=4+\epsilon$. These papers
also showed how to base the security of the hash function on other lattice problems which are potentially harder than
uSVP (e.g., {\sc GapSVP} and {\sc GapSIVP}). In \cite{Micciancio02hash}, Micciancio recently constructed a family of
hash functions with the best known constant $c$ for several important lattice problems (but not for uSVP). In another
paper \cite{Mic02cyclic}, Micciancio improved the efficiency of the hash function by using cyclic lattices. Roughly
speaking, all of these results are based on variations of a method known as Ajtai's iterative step.

\vspace{-1ex}
\subsection*{Our contribution}

The main contribution of this paper is the introduction of Fourier analysis on lattices as an integral part of a
lattice based construction. Fourier analysis was previously used indirectly through transference theorems, i.e.,
theorems that relate properties of a lattice and its dual (see, e.g.,~\cite{CaiAW}). Our constructions are the first to
use Fourier analysis directly.

Our main theorem is a reduction from the $O(n^{1.5})$-uSVP to the problem of distinguishing between two types of
distributions on the segment $[0,1)$. We believe that this theorem will find other uses in the future.

Using the main theorem, we present three results. The main one is a new public key cryptosystem which is based on the
hardness of $O(n^{1.5})$-uSVP. This is a major improvement to the 1996 cryptosystem by Ajtai and Dwork. Its description
is surprising in that it essentially consists only of numbers modulo some large number $N$. Our second result is a
family of collision resistant hash functions whose security is based on the $O(n^{1.5})$-uSVP. In terms of the uSVP,
this improves all the previous results mentioned above. However, previous results were not based only on uSVP and are
therefore incomparable with our result. In addition, ours is the first lattice based hash function whose analysis is
not based on Ajtai's iterative step. The hash function that we consider is simple and is known as the modular subset
sum function\footnote{Previous constructions of hash functions were usually presented as functions on random lattices.
However, most of these results can be easily extended to the modular subset sum function. This was already noted in
Ajtai's original paper~(\cite{Ajtai96}).}. This function already appeared in previous papers; for example, one of the
results in \cite{ImpagliazzoNaor} is an average-case to average-case reduction for the function. The third result is
related to an open question in quantum computation and will be discussed in Section~\ref{section_quantum}.

\vspace{-1ex}
\subsection*{Intuitive overview}

In the following we provide an informal overview of the results in this paper. Many of the details are omitted for the
sake of clarity.

\vskip 10pt
 {\bf Main theorem:} Our main theorem is a reduction from $O(n^{1.5})$-uSVP to the problem of distinguishing between
two types of distributions on $[0,1)$. One distribution is the uniform distribution $U$ while the other $T_h$ is
concentrated around integer multiples of $1/h$ for some {\em unknown} large integer $h \le 2^{O(n^2)}$ (notice that if
we knew $h$ we could easily distinguish between the two). The sharpness of the concentration in this `wavy'
distribution depends on the factor of the uSVP problem. For example, $O(n^{1.5})$-uSVP translates to a concentration of
around $1/n$, that is, the difference between two adjacent peaks is roughly $n$ times the width of a peak (see
Figure~\ref{figure_wavy}). Notice that the reduction is to a worst-case problem in the sense that one has to
distinguish between the uniform distribution and the wavy distribution for all values $h$ in a certain range.
Nevertheless, the wavy distribution has the property that if one distinguishes it from uniform for some small fraction
of $h$ then one can also distinguish it from uniform for all values of $h$. This average-case to worst-case property
will be implicit in our cryptographic applications. In the following we describe the three steps in the proof of the
main theorem.

The first step involves a reduction from the search problem uSVP to a certain decision problem on lattices. Assume that
the shortest vector is $\sum_{i=1}^n a_i v_i$ where $a_i \in \Z$ and $v_1,\ldots,v_n$ is a basis of the lattice. The
decision problem asks whether $p \mid a_i$ where $p$ is some prime number which we choose to be slightly more than
$n^{1.5}$. The reduction is a Cook reduction and the idea is to make the lattice sparser and sparser without losing the
shortest vector. At the end, the lattice is so sparse that we can easily find the shortest vector. For example, when
$p\mid a_i$ we can replace $v_i$ with $p\cdot v_i$ without losing the shortest vector. The actual proof is slightly
more involved as we have to handle cases where $p\nmid a_i$.

The second step is the core of the proof. Here, we reduce the above decision problem to a problem of distinguishing
between two $n$-dimensional distributions. Namely, one distribution is uniform and the other is a `wavy' distribution.
We begin by developing a few lemmas based on a theorem of Banaszczyk. Essentially, this theorem says that if we choose
a `random' lattice point from the dual $L^*$ of a lattice and perturb it by a Gaussian of radius $\sqrt{n}$ then the
distribution obtained can be closely approximated by a function that depends only on points in $L$ (the primal lattice)
that are within distance $\sqrt{n}$ of the origin. We apply this theorem for two types of lattices $L$. The first is a
lattice $L$ where all nonzero vectors are of length more than $\sqrt{n}$. Here we get that the distribution around
points of $L^*$ is determined only by the origin of the primal lattice and is therefore very close to being uniform.
The second type is a lattice with one short vector $u$ of length (say) $1/n$ and all other non-parallel vectors of
length more than $\sqrt{n}$. The distribution that we obtain here is almost uniform on $n-1$ dimensional hyperplanes
orthogonal to $u$. In the direction of $u$ the distribution has peaks of distance $n$ such that the width of each peak
is $1$. The way we use these results is the following. Recall that we are given an $n^{1.5}$-unique lattice and we
should decide whether $p \mid a_i$. We do this by first scaling the lattice so that the length of the shortest vector
is $1/n$ and therefore all non-parallel vectors are of length more than $n^{1.5}/n = \sqrt{n}$. We then multiply $v_i$
by $p$. If $p\mid a_i$ then the shortest vector remains in the lattice and therefore if we take the distribution in the
dual lattice we get a wavy distribution as described above. Otherwise, if $p\nmid a_i$, the shortest vector disappears
and since $p>n^{1.5}$ the resulting lattice has no vectors shorter than $\sqrt{n}$. Therefore, the distribution
obtained in the dual is very close to uniform.

The third and final step consists of `projecting' the $n$-dimensional distributions described above onto a
one-dimensional distribution. Na\"{\i}vely, one can choose a point according to the $n$-dimensional distribution and
project it down to a line. However, this would ruin the original distribution. We would like to project down to a line
but only from tiny areas around the line. This would guarantee that the original distribution is preserved. This,
however, presents a new difficulty: how can one guarantee that a randomly selected point according to the distribution
in $\R^n$ falls close to the line? We solve this by using the fact that the distribution is periodic on the lattice.
Hence it is enough to consider the distribution on the fundamental parallelepiped of the lattice. Now we can draw a
line that passes through the parallelepiped many times and that is therefore `dense' in the $n$-dimensional space
inside the parallelepiped (see Figure~\ref{figure_line_in_ppd}). Projecting the two $n$-dimensional distributions above
will produce either the uniform distribution $U$ or the wavy distribution $T_h$ for some $h$. This completes the
description of the main theorem.

\vskip 10pt
 {\bf Public key cryptosystem:}
Let $N$ be some large integer. The private key consists of a single integer $h$ chosen randomly in the range (say)
$[\sqrt{N}, 2\sqrt{N})$. The public key consists of $m=O(\log N)$ numbers $a_1,\ldots,a_m$ in $\{0,1,\ldots,N-1\}$
which are `close' to integer multiples of $N/h$ (notice that $h$ doesn't necessarily divide $N$). We also include in
the public key an index $i_0 \in [m]$ such that $a_{i_0}$ is close to an {\em odd} multiple of $N/h$. We encrypt one
bit at a time. An encryption of the bit 0 is the sum of a random subset of $\{a_1,\ldots,a_m\}$ modulo $N$. An
encryption of the bit 1 is similar but we add $\floor{a_{i_0}/2}$ to the result. On receiving an encrypted word $w$ we
consider its remainder on division by $N/h$. If it is small, we decrypt 0 and otherwise we decrypt 1. Notice that since
$a_1,\ldots,a_m$ are all close to integer multiples of $N/h$ any encryption of 0 is also close to a multiple of $N/h$
and the decryption is correct. Similarly, since $\floor{a_{i_0}/2}$ is far from a multiple of $N/h$, encryptions of $1$
are also far from multiples of $N/h$ and the decryption is 1.

The following is a rough description of how we establish the security of the cryptosystem. Assume that there exists a
distinguisher $\calA$ that given the public key can distinguish encryptions of 0 from encryptions of 1. In other words,
the difference between the acceptance probabilities $p_0$ on encryptions of 0 and the acceptance probability $p_1$ on
encryptions of 1 is non-negligible. Therefore, if $p_u$ is the acceptance probability on random words $w$, it must be
the case that either $|p_u-p_0|$ or $|p_u-p_1|$ is non-negligible. Assume that the former case holds (the latter case
is similar). Then we construct a distinguisher between the distributions $U$ and $T_h$. Let $R$ be the unknown
distribution on $[0,1)$. We choose $m$ values from $R$, multiply them by $N$ and round the result. Let $a_1,\ldots,a_m$
be the result. We then estimate $\calA$'s acceptance probability when the public key $a_1,\ldots,a_m$ (for simplicity
we ignore $i_0$) is fixed and the word $w$ is chosen randomly as an encryption of 0. This is done by simply calling
$\calA$ many times, each time with a new $w$ computed according to the encryption algorithm. We also estimate $\calA$'s
acceptance probability when $w$ is chosen uniformly from $\{0,1,\ldots,N-1\}$ and not according to the encryption
algorithm. If there is a non-negligible difference between the two estimates, we decide that $R$ is $T_h$ and otherwise
we say that $R$ is $U$. We claim that this distinguishes between $U$ and $T_h$. If $R$ is $U$ then $a_1,\ldots,a_m$ are
uniform in $\{0,1,\ldots,N-1\}$. One can show that this implies that the distribution of encryptions of 0 is very close
to the uniform distribution and therefore $\calA$ (as well as any other algorithm) cannot have different acceptance
probabilities for the two distributions. Otherwise, $R$ is $T_h$ and the distribution that we obtain on
$a_1,\ldots,a_m$ is the same one that is used in the public key algorithm. Therefore, according to our hypothesis,
$\calA$ should have a non-negligible difference between the two cases.

\vskip 10pt
 {\bf A family of collision resistant hash functions:}
We choose $m=O(\log N)$ random numbers $a_1,\ldots,a_m$ uniformly from $\{0,1,\ldots,N-1\}$ and define the hash
function $f(b)=\sum_{i=1}^m b_i a_i ~\mod~ N$ where $b \in \{0,1\}^m$. A collision finding algorithm in this case means
an algorithm $\calA$ that given random $a_1,\ldots, a_m$ finds with non-negligible probability a nonzero vector $b\in
\{-1,0,1\}^m$ such that $\sum b_i a_i \equiv 0 (\mod ~ N)$. Using $\calA$ we show how to build a distinguisher between
$U$ and $T_h$. By trying many values of the form $(1+1/poly(m))^i$ we can have an estimate $\th$ of $h$ up to some
small $1/poly(m)$ error. We would like to use $\th$ to check if the distribution is concentrated around multiples of
$\frac{1}{h}$. Sampling values from the unknown distribution $R$ and reducing modulo $1/\th$ does not help because the
difference between $i/h$ and $i/\th$ is much larger than $1/\th$ for almost all $0\le i < h$ (recall that $h$ is
roughly $\sqrt{N}$ which is exponential in $m$). The idea is to use the collision finding algorithm to create from
$T_h$ a distribution which is also concentrated around the peaks $i\cdot \frac{1}{h}$ but only for $i\le m$.

We sample $m$ values $x_1,\ldots,x_m$ from the unknown distribution $R$. We add small perturbations $y_1,\ldots,y_m$
chosen uniformly in $[0,1/\th)$ to each $x_1,\ldots,x_m$ respectively. We denote the result by $z_1,\ldots,z_m$. Now we
call $\calA$ with $\floor{N\cdot z_1}, \ldots, \floor{N\cdot z_m}$ and we get a subset $S$ such that $\sum_{i\in S} z_i
~\mod~ 1$ is very close to zero. For simplicity assume that it is exactly zero. We then check if $\sum_{i\in S} x_i
~\mod~ 1 = - \sum_{i\in S} y_i ~\mod~ 1$ is close to an integer multiple of $1/\th$. If $R$ is the uniform distribution
on $[0,1)$ then conditioned on any values of $z_1,\ldots,z_m$ the distribution of $y_1,\ldots,y_m$ is still uniform in
$[0,1/\th)$ and hence $\sum_{i\in S} y_i$ is not close to an integer multiple of $1/\th$. If $R$ is $T_h$ then
conditioned on any values of $z_1,\ldots,z_m$, the $x_i$'s are distributed around one or two peaks of $T_h$. Therefore,
$\sum_{i\in S} x_i ~\mod~ 1$ is close to a multiple of $\frac{1}{h}$. Moreover, since the $y_i$'s are at most $1/\th$,
their sum is at most $m/\th$. Since the estimate $\th$ satisfies that for $1\le i\le m$, $i/h$ is very close to
$i/\th$, the distinguisher can reduce $\sum_{i\in S} x_i$ modulo $1/\th$ and see that it is close to a multiple of
$1/\th$, as required.

One last issue that we have to address is that $\calA$ might not find collisions on inputs of the form $\floor{N\cdot
z_1}, \ldots, \floor{N\cdot z_m}$ when $R$ is not the uniform distribution. This is because our assumption was that
$\calA$ finds collisions on inputs chosen uniformly. But if $\calA$ does not find collisions we know that $R$ has to be
$T_h$ and hence we can still distinguish between $U$ and $T_h$.

\vspace{-1ex}
\subsection*{Outline}

In Section~\ref{section_prelim} we list several definitions and some properties of lattices that will be needed in this
paper (for an introduction to lattices see \cite{MicciancioBook}). After defining several distributions in
Section~\ref{section_several_distr} we present the two cryptographic constructions in
Section~\ref{section_crypto_constr}. The main theorem is developed in Section \ref{section_main_theorem}. The analysis
of the public key cryptosystem is in Section~\ref{section_analysis_pke} and that of the hash function is in
Section~\ref{section_analysis_hash}. In Section~\ref{section_quantum} we present a solution to an open problem related
to quantum computation. Several technical claims appear in Appendix~\ref{appendix_technical_claims}.

\section{Preliminaries}\label{section_prelim}

A lattice in $\R^n$ is defined as the set of all integer combinations of $n$ linearly independent vectors. This set of
vectors is known as a basis of the lattice and is not unique. Given a basis $(v_1,\ldots,v_n)$ of a lattice $L$, the
fundamental parallelepiped is defined as
$$\calP(v_1,\ldots,v_n) = \{ \sum_{i=1}^n x_i v_i ~|~ x_i \in [0,1) \}.$$
When the basis is clear from the context we will use the notation $\calP(L)$ instead of $\calP(v_1,\ldots,v_n)$. Note
that a lattice has a different fundamental parallelepiped for each possible basis. We denote by $d(L)$ the volume of
the fundamental parallelepiped of $L$ or equivalently, the determinant of the matrix $B$ whose columns are the basis
vectors of the lattice. The point $x\in \R^n$ reduced modulo the parallelepiped $\calP(v_1,\ldots,v_n)$ is the unique
point $y\in \calP(v_1,\ldots,v_n)$ such that $y-x$ is an integer combination of $v_1,\ldots,v_n$ (see, e.g.,
\cite{Micciancio01hnf}). The dual of a lattice $L$ in $\R^n$, denoted $L^*$, is the set of all vectors $y\in \R^n$ such
that $\ip{x,y}\in \Z$ for all vectors $x\in L$. Similarly, given a basis $(v_1,\ldots,v_n)$ of lattice, we define the
dual basis as the set of vectors $(v_1^*,\ldots,v_n^*)$ such that $\ip{v_i,v_j^*} = \delta_{ij}$ for all $i,j\in [n]$.
Note that if $B=(v_1,\ldots,v_n)$ is the $n\times n$ matrix whose columns are the basis vectors then $(B^T)^{-1}$
contains the dual basis as its columns. From this it follows that $d(L^*)=1/d(L)$.

We say that a lattice is {\em unique} if its shortest vector is strictly shorter than all other non-parallel vectors.
Moreover, a lattice is {\em $f(n)$-unique} if the shortest vector is shorter by a factor of at least $f(n)$ from all
non-parallel vectors. In the shortest vector problem we are interested in finding the shortest vector in a lattice. In
this paper we will be interested in the $f(n)$-unique shortest vector problem ($f(n)$-uSVP) where in addition, we are
promised that the lattice is $f(n)$-unique. Let $\lambda(L)$ denote the length of the shortest nonzero vector in the
lattice $L$. We also denote the shortest vector (or one of the shortest vectors) by $\tau(L)$. Most of the lattices
that appear is this paper are unique lattices and in these cases $\tau(L)$ is unique up to sign.

One particularly useful type of basis is an LLL reduced basis. Such a basis can be found in polynomial time~\cite{LLL}.
Hence, we will often assume without loss of generality that lattices are given by an LLL reduced basis. The properties
of LLL reduced bases that we use are summarized in Claim~\ref{LLL}.

We define a negligible amount as an amount which is asymptotically smaller than $n^{-c}$ for any constant $c>0$. The
parameter $n$ will indicate the input size. Similarly, a non-negligible amount is one which is at least $n^{-c}$ for
some $c>0$. Finally, exponentially small means an expression that is at most $2^{-\Omega(n)}$. We say that an algorithm
$\calA$ with oracle access is a distinguisher between two distributions if its acceptance probability when the oracle
outputs samples of the first distribution and its acceptance probability when the oracle outputs samples of the second
distribution differ by a non-negligible amount. Note that the notion of acceptance is used for convenience. In
addition, an algorithm $\calA$ is said to distinguish between the distribution $T$ and the set of distribution $\calT$
if for any distribution $T' \in \calT$, $\calA$ distinguishes between $T$ and $T'$.

For two continuous random variables $X$ and $Y$ having values in $[0,1)$ with density functions $T_1$ and $T_2$
respectively we define their statistical difference as
$$ \Delta(X,Y) = \frac{1}{2}\int_0^1 |T_1(r) - T_2(r)| dr.$$
A similar definition holds for discrete random variables. One important fact that we use is that the statistical
distance cannot increase by applying a (possibly randomized) function $f$, i.e.,
\begin{equation}\label{eq:statistical_dst}
\Delta(f(X),f(Y)) \le \Delta(X,Y),
\end{equation}
see, e.g., \cite{MicciancioBook}. In particular, this implies that the acceptance probability of any algorithm on
inputs from $X$ differs from its acceptance probability on inputs from $Y$ by at most $\Delta(X,Y)$.

The set $\{1,2,\ldots,n\}$ is denoted by $[n]$. All logarithms are of base 2 unless otherwise specified. We use
$\delta_{ij}$ to denote the Kronecker delta, i.e., 1 if $i=j$ and 0 otherwise. We use $\cc$ to denote an unspecified
constant. That is, whenever $\cc$ appears we can replace it with some universal constant. For example, the expression
$\cc + 7 = \cc$ is true because we can substitute $1$ and $8$ for the constants. Other constants will be denoted by $c$
with a letter as the subscript, e.g., $\C{h}$.

For two real numbers $x,y>0$ we define $x ~\mod~ y$ as $x - \floor{x/y}y$. For $x\in \R$ we define $\round{x}$ as the
integer nearest to $x$ or, in case two such integers exist, the smaller of the two. We also use the notation $\frc(x):=
\abs{x-\round{x}}$, i.e., the distance of a real $x$ to the nearest integer. Notice that for all $x,y\in \R$, $0\le
\frc(x)\le \frac{1}{2}$, $\frc(x) \le \abs{x}$ and $\frc(x+y)\le \frc(x)+\frc(y)$.

Recall that the {\em normal distribution} with mean 0 and variance $\sigma^2$ is the distribution on $\R$ given by the
density function $\frac{1}{\sqrt{2\pi}\cdot \sigma} e^{-\frac{1}{2}(\frac{x}{\sigma})^2}$. Also recall that the sum of
two normal variables with mean 0 and variances $\sigma_1^2$ and $\sigma_2^2$ is a normal variable with mean 0 and
variance $\sigma_1^2+\sigma_2^2$. A simple tail bound on the normal distribution appears in Claim~\ref{tail_bound}. The
{\em Gaussian} distribution is a distribution on $\R^n$ obtained by taking $n$ independent identically distributed
normal random variables as the coordinates. We define the {\em standard Gaussian} distribution as the distribution
obtained when each of the normal random variables has standard deviation $1/\sqrt{2\pi}$. In other words, a standard
Gaussian distribution is given by the density function $e^{-\pi \|x\|^2}$ on $\R^n$.

For clarity, we present some of our reductions in a model which allows operations on real numbers. It is possible to
modify them in a straightforward way so that they operate in a model that approximates real numbers up to an error of
$2^{-n^c}$ for arbitrary large constant $c$ in time polynomial in $n$. Therefore, if we say that two continuous
distributions on $[0,1)$ are indistinguishable (in the real model) then for any $c>0$ discretizing the distributions up
to error $2^{-n^c}$ for any $c$ yields two indistinguishable distributions.

\subsection{Several Distributions}\label{section_several_distr}

\begin{figure}[h]
\begin{center}
$\begin{array}{c@{\hspace{3mm}}c@{\hspace{3mm}}c}
    \epsfxsize=2in \epsffile{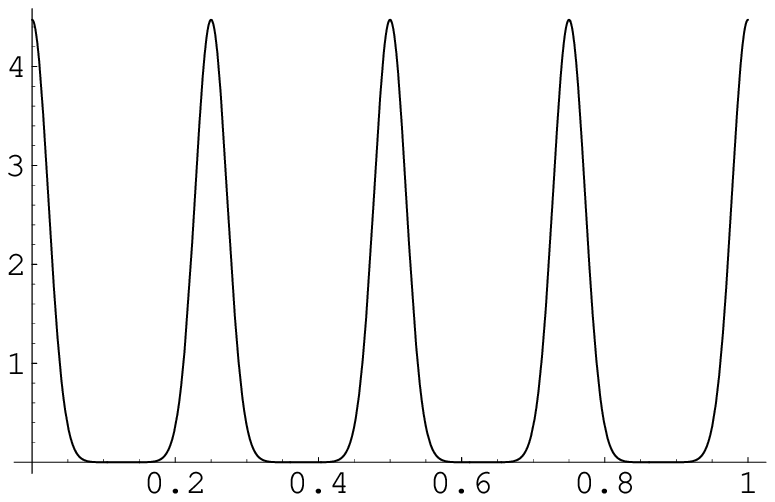} &
    \epsfxsize=2in \epsffile{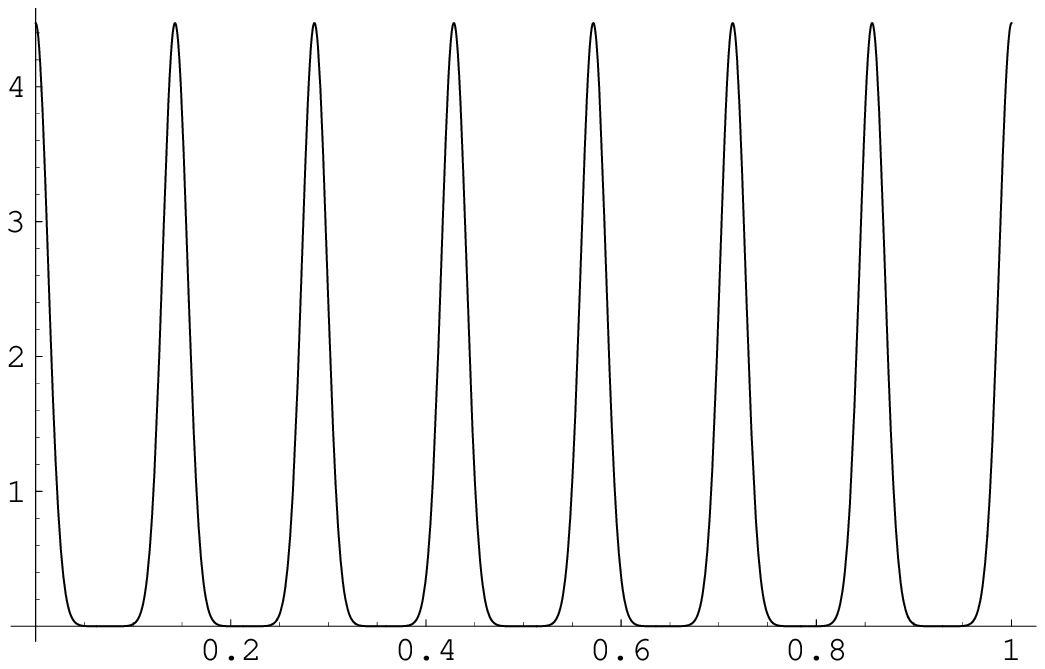} &
    \epsfxsize=2in \epsffile{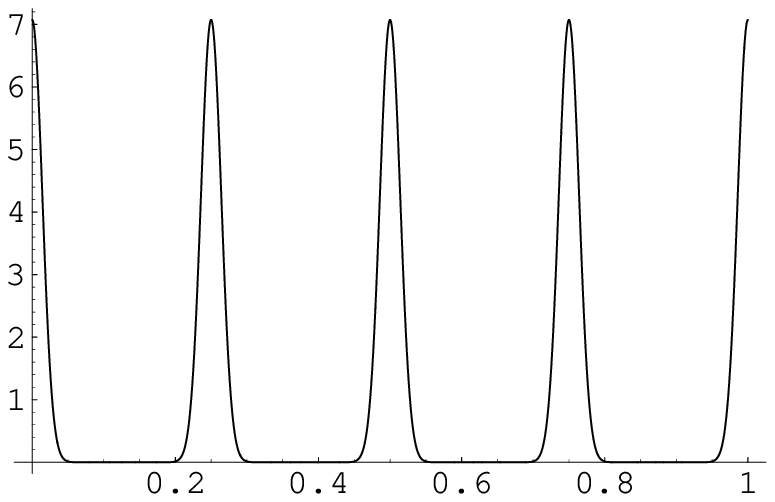}
\end{array}$
\end{center}
\caption{$T_{4,0.05}$, $T_{7,0.05}$ and $T_{4,0.02}$} \label{figure_wavy}
\end{figure}

We define several useful distributions on the segment $[0,1)$. The distribution $U$ is simply the uniform distribution
with the density function $U(r)=1$. For $\beta\in \R^+$ the distribution $Q_\beta$ is a normal distribution with mean
$0$ and variance $\frac{\beta}{2\pi}$ reduced modulo $1$ (i.e., a periodization of the normal distribution):
$$Q_\beta(r) =  \sum_{k=-\infty}^{\infty} \frac{1}{\sqrt{\beta}} e^{-\frac{\pi}{\beta}(r-k)^2}.$$
Clearly, one can efficiently sample from $Q_\beta$ by sampling a normal variable and reducing the result modulo 1.
Another distribution is $T_{h,\beta}$ where $h \in \N$ and $\beta \in \R^+$ (see Figure~\ref{figure_wavy}). Its density
function is defined as
$$T_{h,\beta}(r) =  Q_\beta(rh~\mod~1) = \sum_{k=-\infty}^{\infty} \frac{1}{\sqrt{\beta}} e^{-\frac{\pi}{\beta}(rh-k)^2}.$$
By adding a normalization factor we can extend the definition of $T_{h,\beta}$ to non-integer $h$. So in general,
$$T_{h,\beta}(r) =  \frac{1}{\int_{0}^1 Q_\beta(xh~\mod~1) dx} Q_\beta(rh~\mod~1) .$$
For a real $h> 0$, choosing a value $z\in [0,1)$ according to $T_{h,\beta}$ can be done as follows. First choose a
value $x\in \{0,1,\ldots,\ceil{h}-1\}$ and then choose a value $y$ according to $Q_\beta$. If $\frac{x+y}{h}<1$ then
return it as the result. Otherwise, repeat the process again. It is easy to see that the distribution obtained is
indeed $T_{h,\beta}$ and that the process is efficient for (say) $h\ge 1$.

We also define the following set of distributions:
$$ \calT_{n,g} := \{ T_{h,\beta} ~|~ h\in \N,~ h \le 2^{\C{h} n^2},~\beta\in [\frac{n}{g^2}, 4\frac{n}{g^2}) \} $$
where $\C{h}$ is a constant specified in Lemma~\ref{lemma_one_dim_distr}.

\section{Cryptographic Constructions}\label{section_crypto_constr}

For a security parameter $n$, let $N$ be $2^{\C{N} n^2}$ and let $m$ be $\C{m} n^2$ where $\C{N}$ and $\C{m}$ are two
constants which will be specified later. Let $\gamma(n)=\omega(n \sqrt{\log n})$, i.e., any function that satisfies
$\frac{\gamma(n)}{n \sqrt{\log n}} \rightarrow \infty$ as $n$ goes to infinity. The smaller the function, the better
the security guarantee becomes. For concreteness, one can choose $\gamma(n) = n \log n$. We also assume that $\gamma(n)
\le n^{\C{\gamma}}$ for some constant $\C{\gamma}>0$.

\vskip 0.4cm
 {\setlength{\parindent}{2mm}\bf Public Key Encryption}
 \begin{itemize}
 \item
 {\bf Private key:} Let $H=\{h\in [\sqrt{N}, 2\sqrt{N}) ~|~ \frc(h) < \frac{1}{16 m} \}$. Choose $h\in H$ uniformly at random.
                Let $d$ denote $\frac{N}{h}$. The private key is the number $h$.
 \item
 {\bf Public Key:} Choose $\beta\in [4\frac{1}{(\gamma(n))^2},8\frac{1}{(\gamma(n))^2})$ uniformly
                at random. We choose $m$ values $z_1,\ldots,z_m$ from $T_{h, \beta}$ by choosing $x_1,\ldots,x_m$ and
                $y_1,\ldots,y_m$ as described in Section~\ref{section_several_distr}. Let $i_0$ be an index such that $x_{i_0}$ is odd
                (such an $i_0$ exists with probability exponentially close to $1$). For $i \in [m]$, let $a_i$ denote
                $\floor{N\cdot z_i}$. The public key is $(a_1,\ldots,a_m,i_0)$.
 \item
 {\bf Encryption:} In order to encrypt a bit we choose a random subset $S$ of $[m]$. The encryption is $\sum_{i\in S} a_i ~\mod~
 N$ if the bit is 0 and $\sum_{i\in S} a_i + \floor{\frac{a_{i_0}}{2}}~\mod~ N$ if the bit is 1.
 \item
 {\bf Decryption:} On receiving $w\in \{0,\ldots,N-1\}$ we decrypt $0$ if $\frc(\frac{w}{d}) < \frac{1}{4}$ and $1$ otherwise.
 \end{itemize}

\vskip 0.4cm
 {\setlength{\parindent}{2mm}\bf A Family of Collision Resistant Hash Functions}
 \begin{itemize}
 \item Choose $m$ numbers $a_1,\ldots,a_m$ uniformly in $\{0,1,\ldots,N-1\}$. The function $f:\{0,1\}^m \rightarrow
 \{0,1,\ldots,N-1\}$ is defined as:
 $$f(b) = \sum_{i=1}^m b_i a_i ~\mod~ N.$$
 \end{itemize}

Notice that if $c_m > c_N$ then $f$ indeed compresses the size of the input and collisions are guaranteed to exist.

\section{Main Theorem}\label{section_main_theorem}

In this section we present a reduction from $g(n)$-uSVP to the problem of distinguishing between two types of
distributions on $[0,1)$.

\begin{theorem}\label{thm_main_theorem}
Let $g(n)$ be any function such that $4\sqrt{n} \le g(n) \le poly(n)$ and let $\C{h}$ the constant specified in
Lemma~\ref{lemma_one_dim_distr}. If there exists a distinguisher between $U$ and $\calT_{n,g(n)}$ then there exists a
solution to $g(n)$-uSVP.
\end{theorem}
\begin{proof}
Let $p(n)$ be a prime larger than $g(n)$ and at most (say) $2g(n)$. We can now apply Lemmas
\ref{lemma_search_to_decision}, \ref{lemma_two_indis} and \ref{lemma_one_dim_distr} in order to obtain the theorem.
\end{proof}

\subsection{Reduction to a Decision Problem}\label{section_decision_problem}

We reduce the SVP to the following decision problem:

 \vskip 0.4cm
 {\setlength{\parindent}{2mm}\bf Decision SVP with parameter $p$ ($dSVP_p$)}
 \begin{itemize}
 \item
 {\bf Input:} An arbitrary basis $(v_1,\ldots, v_n)$ of a unique lattice $L$ and a number $\alpha$ such that
   $\lambda(L) < \alpha \le 2 \lambda(L)$ and let $\tau(L)=\sum_{i=1}^n a_i v_i$ be the coefficients of
   the shortest vector.
 \item
 {\bf Output:} YES if $p$ divides $a_1$, NO otherwise.
 \end{itemize}

\begin{lemma}\label{lemma_search_to_decision}
Let $p=p(n)>2$ be a prime number which is at most polynomial in $n$.\footnote{The result holds for the case $p=2$ as
well with some technical differences.} There exists a reduction from finding the shortest vector in a unique lattice
$L$ to $dSVP_p$.\footnote{One can guarantee the uniqueness of the shortest vector in any lattice by adding tiny
perturbations to the basis vectors. Therefore, the assumption that $L$ is unique can be avoided.} Moreover, if $L$ is
an $f(n)$-unique lattice then all the calls to $dSVP$ are also with an $f(n)$-unique lattice.
\end{lemma}
\begin{proof}
It is convenient to have a bound on the coefficients of the shortest vector. So we assume, without loss of generality,
that we are given a basis $(v_1,\ldots, v_n)$ of $L$ which is LLL reduced. Hence, by Claim~\ref{LLL}, we get that the
coefficients of the shortest vector satisfy $|a_i| \le 2^{2n}$ and $\frac{\|v_1\|}{2^{n}} \le \lambda(L) \le \|v_1\|$.
These are the only properties that we need from the basis and in fact, other bases used throughout this proof will not
necessarily be LLL reduced. In the following we describe a procedure $\calB(\alpha)$ that finds the shortest vector
given an estimate $\alpha$ which satisfies $\lambda(L) < \alpha \le 2\lambda(L)$. We apply the procedure $n$ times with
$\alpha=2^{j-n}\cdot \|v_1\|$ for $j=1,2,\ldots, n+1$. Notice that when we call $\calB$ with the wrong value of
$\alpha$ it can error by either outputting a non-lattice vector or a lattice vector which is longer than the shortest
vector. We can easily ignore these errors by checking that the returned vector is a lattice vector and then take the
shortest one. Therefore, it is sufficient to show that when $\alpha$ satisfies $\lambda(L) < \alpha \le 2\lambda(L)$,
$\calB(\alpha)$ returns the shortest vector. Clearly, one can modify the $dSVP$ so that it finds whether $p~|~a_i$ for
any $i\in [n]$ (and not just $i=1$) by simply changing the order of the vectors in the basis given to the $dSVP$.

The procedure $\calB$ is based on changes to the basis $(v_1,\ldots, v_n)$. Throughout the procedure we maintain the
invariant that the lattice spanned by the current basis is a sublattice of the original lattice and that the shortest
vector is unchanged. Notice that this implies that if the original lattice is an $f(n)$-unique lattice then all
intermediate lattices are also $f(n)$-unique and hence all the calls to $dSVP$ are with an $f(n)$-unique lattice, as
required. In addition, since the shortest vector is unchanged, the estimate $\alpha$ can be used whenever we call the
dSVP with an intermediate lattice. The changes to the basis are meant to decrease the coefficients of the shortest
vector. We let $a_1,\ldots,a_n$ denote the coefficients of the shortest vector according to the current basis. We will
show that when the procedure ends all the coefficients of the shortest vector are zero except $a_i$ for some $i\in
[n]$. This implies that the shortest vector is $v_i$. In the following we describe a routine $\calC$ that will later be
used in $\calB$.

The routine $\calC(i,j)$ where $i,j\in [n]$ applies a sequence of changes to the basis. Only the vectors $v_i$ and
$v_j$ in the basis are modified. When the routine finishes it returns the new basis and a bit. If the bit is zero then
we are guaranteed that the coefficient $a_i$ of the shortest vector in the new basis is zero. Otherwise, the bit is one
and we are guaranteed that $|a_j| \le \frac{1}{2} |a_i|$ and that $a_i$ is nonzero. In any case, the value of $|a_i|$
does not increase by $\calC(i,j)$.

The routine is composed of the following two steps. In the first step we replace $v_i$ with $p\cdot v_i$ as long as the
$dSVP$ says that $p \mid a_i$ and not more than $2n$ times. By multiplying $v_i$ by $p$ when $p\mid a_i$, we obtain a
sublattice that still contains the same shortest vector. The coefficient $a_i$ decreases by a factor of $p$. Since we
began with $|a_i| < 2^{2n}$, if this happens $2n$ times then $a_i = 0$ and therefore in this case we return the current
lattice and output a zero bit. Otherwise, we are guaranteed that in the current lattice $p\nmid a_i$.

In the second step we consider $p$ different bases where $v_i$ is replaced with one of $v_i-\frac{p-1}{2}
v_j,\ldots,v_i-v_j,v_i,v_i+v_j,\ldots, v_i+\frac{p-1}{2} v_j$. Notice that all $p$ bases span the same lattice. Also
note that the coefficient $a_j$ changes to $a_j+\frac{p-1}{2} a_i,\ldots,a_j+a_i,a_j,a_j-a_i,\ldots, a_j-\frac{p-1}{2}
a_i$ respectively while all other coefficients remain the same. Since $p\nmid a_i$, one of the bases must satisfy that
$p \mid a_j$ and we can find it by calling $dSVP_p$. We choose that basis and then multiply $v_j$ by $p$. We repeat the
above steps (of choosing one of the $p$ bases and multiplying by $p$) $2n$ times and then output the resulting lattice
with the bit one. With each step, the new $|a_j|$ becomes at most $(\frac{p-1}{2} |a_i| + |a_j| )/p =
(\frac{1}{2}-\frac{1}{2p}) |a_i| + \frac{|a_j|}{p}$. Hence, after $2n$ applications, the new $|a_j|$ is at most
$(\frac{1}{2}-\frac{1}{2p})(1+\frac{1}{p}+\ldots+\frac{1}{p^{2n-1}})|a_i| + \frac{|a_j|}{p^{2n}} <
 \frac{1}{2}|a_i| + \frac{|a_j|}{p^{2n}}$ and since $a_j$ is integer this implies $|a_j| \le \frac{1}{2}|a_i|$. This
completes the description of $\calC$. It is easy to check that all the numbers involved have a polynomial size
representation and therefore $\calC$ runs in polynomial time.

The procedure $\calB$ works by maintaining a set $Z$ of possibly non-zero coefficients which is initially set to $[n]$.
As long as $|Z|\ge 2$ we perform the following operations. Assume without loss of generality that $1,2\in Z$. We
alternatively call $\calC(1,2)$ and $\calC(2,1)$ until the bit returned in one of the calls is zero. This indicates
that one of the coefficients is zero (either $a_1$ or $a_2$ depending on which call returns the zero bit) and we remove
it from the set $Z$. In order to show that the procedure runs in polynomial time, it is enough to show that an element
is removed from $Z$ after at most a polynomial number of steps. Notice that after each pair of calls to $\calC$ that
returned the bit one $|a_1|$ decreases by a factor of at least 4. Therefore, after at most $2n$ calls to $\calC$, $a_1$
becomes zero and $\calC(1,2)$ must return the bit zero.
\end{proof}

Although not used in this paper, the following is an immediate corollary of the above lemma and might be of independent
interest. Basically, it is a reduction from the search SVP to the decision SVP for unique lattices. It is still an open
question whether a similar result holds for SVP on general lattices.
\begin{corollary}\label{corollary_search_to_decision}
For any prime $p=p(n)< poly(n)$ larger than 2 and any $f(n)\ge 1$, finding the shortest vector in an $p(n)f(n)$-unique
lattice can be reduced to the following gap problem: given $d$ and an $f(n)$-unique lattice, decide whether the length
of the shortest vector is at most $d$ or more than $\sqrt{p(n)}\cdot d$.
\end{corollary}
\begin{proof}
According to Lemma~\ref{lemma_search_to_decision} it is enough to describe a solution to $dSVP_p$ on $p\cdot
f(n)$-unique lattices. Say we are given the lattice $L$ with the basis $(v_1,\ldots,v_n)$. By using the gap problem we
can approximate $\lambda(L)$ and $\lambda(L')$ up to a factor $\sqrt{p}$ where $L'$ is the lattice spanned by $(p
v_1,v_2,\ldots,v_n)$. Notice that since $p\cdot \tau(L) \in L'$, $L'$ is an $f(n)$-unique lattice, as required. Say
that $\lambda(L) \in [d_1,\sqrt{p} d_1]$ and $\lambda(L')\in [d_2,\sqrt{p} d_2]$. If $p \mid a_1$ then both lattices
contain the same shortest vector and therefore the two ranges intersect. Otherwise, it is easy to see that $\tau(L') =
p\cdot \tau(L)$ and therefore the two ranges do not intersect.
\end{proof}

\subsection{Gaussian Distributions on Lattices}\label{section_gaussian_distributions}

Let $B_n$ denote the Euclidean unit ball and define $\rho(A)$ as $\sum_{x\in A} e^{-\pi \|x\|^2}$. The following lemma
by Banaszczyk says that in any lattice $L$ the contribution to $\rho(L)$ from points of distance more than $\sqrt{n}$
is negligible.
\begin{lemma}[\cite{Banaszczyk}, Lemma 1.5(i) with $c=1$]\label{bana}
For any lattice $L$, $\rho(L-\sqrt{n}B_n) < 2^{-\Omega(n)} \rho(L)$.
\end{lemma}

The proof of this lemma is not straightforward; a somewhat easier proof can be found in \v{S}tefankovi\v{c}'s thesis
\cite{StefankovicThesis}. A simple corollary of this lemma is that $\rho(L) < \rho(L \cap
\sqrt{n}B_n)/(1-2^{-\Omega(n)})$. We will also use the following formulation of the Poisson summation formula:
\begin{lemma}[\cite{Banaszczyk}, Lemma 1.1(i) with $a=\pi$, $b=1$, $y=0$]\label{psf}
For any lattice $L$ and any vector $y\in \R^n$, $\rho(L^*+y) = d(L)\cdot \sum_{x\in L}e^{2\pi i \ip{x,y}}\rho(\{x\})$.
\end{lemma}

For a given lattice $L$, we consider the distribution obtained by sampling a standard Gaussian centered around the
origin and reducing it modulo the fundamental parallelepiped $\calP(L^*)$. Equivalently, we consider the following
density function defined on $\calP(L^*)$:
$$D_{L^*}(x) = \rho(L^*+x).$$
Intuitively, we can think of $D_{L^*}$ as taking Gaussian distributions around `all' points of $L^*$. Since this
distribution is periodic in $\R^n$ with period $\calP(L^*)$, we simplify the analysis by choosing $D_{L^*}$ to be a
restriction of the distribution to $\calP(L^*)$. In this section we present good approximations to $D_{L^*}$ for two
types of lattices $L$.

\begin{lemma}\label{gaussian_uniform}
Let $L$ be a lattice in which all non-zero vectors are of length more than $\sqrt{n}$ and let
$U_{L^*}(x)=\frac{1}{d(L^*)}=d(L)$ be the uniform density function on $\calP(L^*)$. Then, $\Delta(D_{L^*}, U_{L^*}) <
2^{-\Omega(n)}$.
\end{lemma}
\begin{proof}
For any $y \in \R^n$,
\begin{eqnarray*}
 &&|1 - \sum_{x\in L} e^{2\pi i \ip{x,y}} \rho (\{x\}) | \le
   \sum_{x\in L\setminus \sqrt{n} B_n} \rho(\{x\}) = \\
 && \rho(L\setminus \sqrt{n}B_n) \stackrel{\{1\}}{<} 2^{-\Omega(n)} \rho(L) \stackrel{\{2\}}{<}
 2^{-\Omega(n)} \frac{\rho(L\cap \sqrt{n}B_n)}{1-2^{-\Omega(n)}}  \le
 2^{-\Omega(n)}
\end{eqnarray*}
where $\{1\}$ and $\{2\}$ are due to Lemma~\ref{bana} and the last inequality holds because $\rho(L\cap \sqrt{n}B_n) =
1$. Multiplying by $d(L)$ and using Lemma~\ref{psf} we get,
 $$ |d(L) - \rho(L^*+y)|< 2^{-\Omega(n)}d(L).$$
We conclude the proof by integrating over $\calP(L^*)$,
 $$ \Delta(D_{L^*}, U_{L^*}) < 2^{-\Omega(n)}.$$
\end{proof}

For any vector $v\in L$ define the density function $T_{L^*,v}$ on $\calP(L^*)$ as
$$T_{L^*,v}(x) = \frac{d(L)}{\|v\|}\sum_{k\in \Z} e^{-\pi (\frac{k + \ip{v,x}}{\|v\|})^2}.$$
According to Claim~\ref{t_density_function} it is indeed a density function.

\begin{lemma}\label{gaussian_wavy}
Let $L$ be a lattice with a shortest vector $u$ in which all vectors not parallel to $u$ are of length more than
$\sqrt{n}$. Then, $\Delta(D_{L^*}, T_{L^*,u}) < 2^{-\Omega(n)} (1+\frac{1}{\|u\|})$. In particular, if $\|u\| \ge
\frac{1}{n^c}$ for some $c>0$ then $\Delta(D_{L^*}, T_{L^*,u}) < 2^{-\Omega(n)}$.
\end{lemma}
\begin{proof}
For any $y \in \R^n$,
\begin{eqnarray*}
 &&\abs{ \sum_{x\in L} e^{2\pi i \ip{x,y}} \rho (\{x\}) - \sum_{k \in \Z} e^{2\pi i k \ip{u,y}} \rho(\{k u\}) }< \\
 &&\sum_{x\in L\setminus \sqrt{n} B_n} \rho(\{x\}) = \rho(L\setminus \sqrt{n}B_n) \stackrel{\{1\}}{<} 2^{-\Omega(n)} \rho(L)
 \stackrel{\{2\}}{\le} 2^{-\Omega(n)} \frac{1}{1-2^{-\Omega(n)}} \rho(\{ku | k\in \Z\}) \le\\
 &&2^{-\Omega(n)} \rho(\{ku | k\in \Z\}) \stackrel{\{3\}}{\le} 2^{-\Omega(n)} (1 + \frac{1}{\|u\|})
\end{eqnarray*}
where $\{1\}$ and $\{2\}$ are due to Lemma~\ref{bana} and $\{3\}$ is due to Claim~\ref{sum_exp} with $x=0$. By
multiplying by $d(L)$ we get
\begin{equation}\label{rhodist}
 |\rho(L^*+y) - d(L) \sum_{k \in \Z} e^{2\pi i k \ip{u,y}} \rho(\{k u\})|< 2^{-\Omega(n)} (1 + \frac{1}{\|u\|}) \cdot d(L).
\end{equation}
Consider the one dimensional lattice $M$ spanned by the number $\|u\|$. Clearly, the lattice $M^*$ is spanned by the
number $\frac{1}{\|u\|}$. According to Lemma~\ref{psf} for any $a\in \R$,
$$\rho(M^*+a) = d(M)\sum_{b\in M}e^{2\pi i ab}\rho(\{b\}) = \|u\| \sum_{k\in \Z}e^{2\pi i k a\|u\|}\rho(\{ku\}).$$
Therefore, taking $a= \ip{u,y}/\|u\|$,
\begin{eqnarray*}
 &&d(L) \sum_{k \in \Z} e^{2\pi i k \ip{u,y}} \rho(\{k u\}) =
    \frac{d(L)}{\|u\|}\rho(M^*+\frac{\ip{u,y}}{\|u\|}) =
    T_{L^*,u}(y)
\end{eqnarray*}
We conclude the proof by integrating (\ref{rhodist}) over $\calP(L^*)$:
 $$\Delta(D_{L^*}, T_{L^*,u}) < 2^{-\Omega(n)} (1 + \frac{1}{\|u\|}).$$
\end{proof}

\subsection{Two Indistinguishable Distributions}\label{section_two_indistinguishable}

\begin{lemma}\label{lemma_two_indis}
Let $g(n)<p(n)$ be such that $p(n)$ is a prime and both are at most polynomial in $n$. Solving $dSVP_{p(n)}$ on
$g(n)$-unique lattices can be reduced to the problem of distinguishing between $U_{L^*}$ and $T_{L^*,\tau(L)}$ where
$L$ is given as an $LLL$ reduced basis and $\lambda(L)\in [\frac{\sqrt{n}}{g(n)},\frac{2\sqrt{n}}{g(n)})$.
\end{lemma}
\begin{proof}
We are given a basis $(v_1,\ldots,v_n)$ of a $g(n)$-unique lattice $L$ and a number $\alpha$ such that $\lambda(L) <
\alpha \le 2 \lambda(L)$. Let $L'$ be the lattice $L$ scaled by a factor $\frac{2\sqrt{n}}{\alpha\cdot g(n)}$, i.e.,
the lattice spanned by the basis $(v_1',\ldots,v_n') := \frac{2\sqrt{n}}{\alpha\cdot g(n)}(v_1,\ldots,v_n)$. Notice
that in $L'$ the shortest vector $\tau(L')=\sum_{i=1}^n a_i v_i'$ is of length in
$[\frac{\sqrt{n}}{g(n)},\frac{2\sqrt{n}}{g(n)})$ and any vector not parallel to $\tau(L')$ is of length at least
$g(n)\cdot \frac{\sqrt{n}}{g(n)} = \sqrt{n}$. Now, let $M$ be the lattice spanned by the basis
$(p(n)v_1',v_2',\ldots,v_n')$. If $p(n) \mid a_1$ then $\tau(M)=\tau(L')$ and therefore its length is in
$[\frac{\sqrt{n}}{g(n)},\frac{2\sqrt{n}}{g(n)})$. Also, since $M\subseteq L'$, any vector in $M$ not parallel to
$\tau(M)$ is of length at least $\sqrt{n}$. If $p(n) \nmid a_1$ then the shortest multiple of $\tau(L')$ that is
contained in $M$ is $p(n) \cdot \tau(L')$ whose length is at least $p(n) \cdot \frac{\sqrt{n}}{g(n)}> \sqrt{n}$. Hence,
in this case all non-zero vectors are of length at least $\sqrt{n}$.

Clearly, we can take an LLL reduced basis of the lattice $M$ without changing the properties of the lattice described
above. Now consider the distribution $D_{M^*}$. We can efficiently sample from it by sampling a Gaussian centered
around the origin with standard deviation $\frac{1}{\sqrt{2\pi}}$ and reducing it modulo $\calP(M^*)$. According to
Lemma~\ref{gaussian_wavy}, if $p(n)\mid a_1$ then the distribution is exponentially close to $T_{M^*,\tau(M)}$. On the
other hand, if $p(n)\nmid a_1$, Lemma~\ref{gaussian_uniform} says that the distribution is exponentially close to the
uniform distribution $U_{M^*}$. Therefore, we can decide with non-negligible probability if $p(n)\mid a_1$ by calling
an algorithm that distinguishes between $T_{M^*,\tau(M)}$ and $U_{M^*}$. The error probability can be made
exponentially small by calling the algorithm a polynomial number of times and taking the majority.
\end{proof}

\subsection{One Dimensional Distributions}\label{section_one_dimensional}

\begin{lemma}\label{lemma_one_dim_distr}
There exists a constant $\C{h}$ such that for any $g(n) \ge 4\sqrt{n}$, $g(n) \le poly(n)$, the problem of
distinguishing between $U_{L^*}$ and $T_{L^*,\tau(L)}$ for a lattice $L$ given as an LLL reduced basis for which
$\lambda(L) \in [\frac{\sqrt{n}}{g(n)},\frac{2\sqrt{n}}{g(n)})$ can be reduced to the problem of distinguishing between
$U$ and $\calT_{n,g(n)}$.
\end{lemma}
\begin{proof}
Let $v_1,\ldots,v_n$ denote the $LLL$ reduced basis of $L$ and let $v^*_1,\ldots,v^*_n$ be the dual basis of $L^*$,
i.e., a basis of $L^*$ such that $\ip{v_i,v^*_j}=\delta_{ij}$. For some large integer $K$ to be chosen later, consider
a function $f$ which maps a vector $v=\sum_{i=1}^n a_i v^*_i$ in $\calP(L^*)$ to $\frac{\floor{K a_1}}{K} +
\frac{\floor{K a_2}}{K^2} + \ldots + \frac{\floor{K a_{n-1}}}{K^{n-1}} + \frac{a_n}{K^n} \in [0,1)$. An equivalent way
to describe $f$ is the following. For a real $r\in [0,1)$ let $r_1,\ldots,r_{n-1}\in
\{0,\frac{1}{K},\ldots,\frac{K-1}{K}\}$ and $r_n\in [0,1)$ be the unique numbers such that
$r=r_1+\frac{1}{K}r_2+\ldots+\frac{1}{K^{n-2}}r_{n-1}+\frac{1}{K^{n-1}}r_n$. The set of points that are mapped to $r$
is given by
$$ S(r):=\{ \sum_{i=1}^n a_i v^*_i ~|~ \forall i \in [n-1]~ a_i \in [r_i, r_i+\frac{1}{K}] ~\mbox{and}~ a_n = r_n\}.$$
Hence, $S(r)$ is an $n-1$ dimensional parallelepiped whose diameter is at most
$\frac{1}{K}\sum_{i=1}^{n-1}\norm{v^*_i}$. Let $w$ denote the vector $v^*_1+K v^*_2+\ldots+K^{n-1}v^*_n$. Then, it is
easy to see that for any $r\in [0,1)$ the point $rw$ reduced modulo $\calP(L^*)$ is contained in $S(r)$. The line
connecting the origin with $w$ reduced modulo $\calP(L^*)$ goes through the parallelepiped $K^{n-1}$ times. The mapping
$f$ essentially takes each point in $\calP(L^*)$ to a nearby point on the line (see Figure~\ref{figure_line_in_ppd}).

 \begin{figure}[t]
    \begin{center}
    $\begin{array}{c@{\hspace{10mm}}c}
    \epsfxsize=1.6in \epsffile{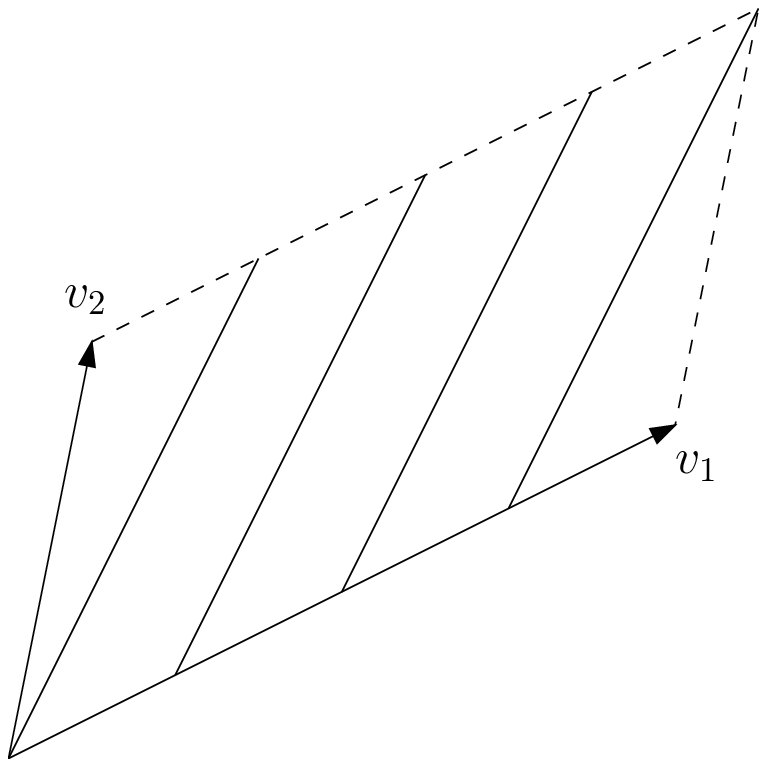} &
    \epsfxsize=1.8in \epsffile{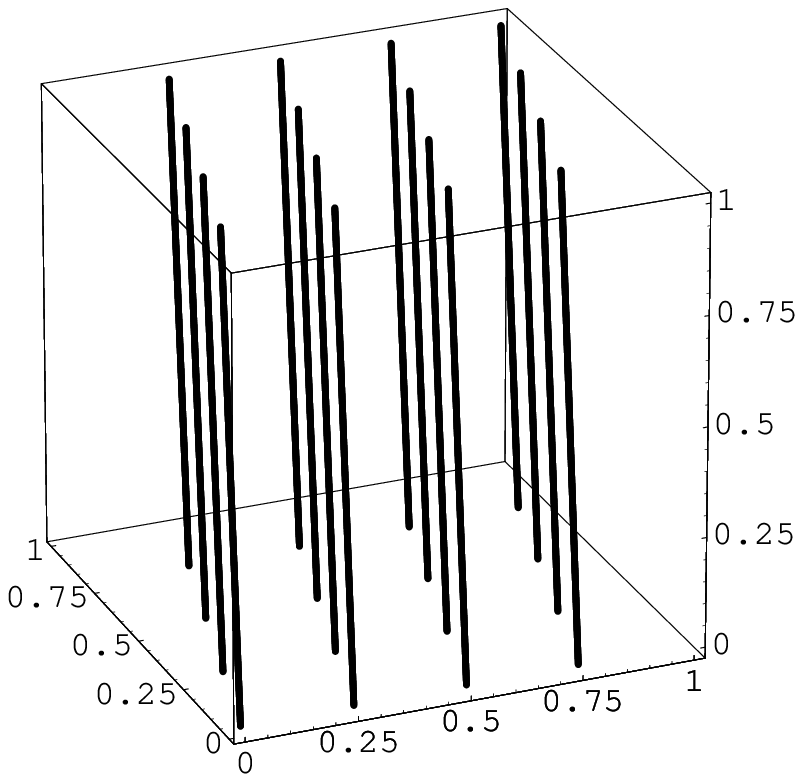}
    \end{array}$
    \end{center}
    \caption{The line connecting the origin to $w$ with $K=4$ in two dimensions with $\calP(v_1,v_2)$ and in three dimensions with the unit cube.}
    \label{figure_line_in_ppd}
 \end{figure}

The reduction works by sampling a point from the given distribution on $\calP(L^*)$ and applying $f$, thereby obtaining
a distribution on $[0,1)$. Notice that $f$ can be computed efficiently. Clearly, by starting from a uniform
distribution on $\calP(L^*)$ we obtain the uniform distribution on $[0,1)$. Hence, it is enough to consider
$T_{L^*,\tau(L)}$. The distribution that we get on $[0,1)$ is given by:
$$ T_1(r) := \frac{d(L^*)}{\vol(S(r))} \int_{S(r)} T_{L^*,\tau(L)}(x) dx$$
which is $d(L^*)$ times the average of $T_{L^*,\tau(L)}$ over $S(r)$. We claim that by choosing $K$ to be large enough
this average is very close to its value in $rw \in S(r)$. More formally, we claim that $T_1(r)$ is exponentially close
to
\begin{eqnarray*}
&& T_{|\ip{\tau(L),w}|,\lambda(L)^2}(r) =
   \frac{1}{\lambda(L)}\sum_{k\in \Z} e^{-\pi (\frac{k + r\ip{\tau(L),w}}{\lambda(L)})^2} =
   d(L^*) T_{L^*,\tau(L)}(rw)
\end{eqnarray*}
where in the first equality we used the fact that $\ip{\tau(L),w}$ is integer and that the function does not change if
we change the sign of $\ip{\tau(L),w}$.

By using the mean value theorem we get that for any $r\in [0,1)$ the difference between the maximum and the minimum
values of $T_{L^*,\tau(L)}$ over $S(r)$ is at most:
\begin{eqnarray*}
 &&\diam(S(x))\cdot \max_x \frac{d}{dx}(\frac{d(L)}{\lambda(L)}\sum_{k\in \Z} e^{-\pi (\frac{k + \lambda(L) x}{\lambda(L)})^2}) \le
 \cc\cdot\frac{1}{K} \sum_{i=1}^{n-1}\norm{v^*_i} \cdot \frac{d(L)}{\lambda(L)}
\end{eqnarray*}
where the inequality is due to Claim~\ref{claim_normal_bound} and the assumption that $\lambda(L) \le
\frac{2\sqrt{n}}{g(n)}\le \frac{1}{2}$. Hence, using Claim~\ref{LLL},
\begin{eqnarray*}
 \forall r\in [0,1)~~\abs{T_1(r) - T_{|\ip{\tau(L),w}|,\lambda(L)^2}(r)} &\le& d(L^*)\cdot \cc\cdot\frac{1}{K} \sum_{i=1}^{n-1}\norm{v^*_i} \cdot \frac{d(L)}{\lambda(L)} =
  \cc\cdot\frac{1}{K} \sum_{i=1}^{n-1}\norm{v^*_i} \cdot \frac{1}{\lambda(L)} \\
 &\le& \cc \cdot \frac{1}{K} \cdot n \cdot \frac{\sqrt{n}}{\lambda(L)} \cdot 2^{2n} \cdot \frac{1}{\lambda(L)} \le
   \cc \cdot \frac{1}{K} \cdot 2^{2n} \cdot poly(n)
\end{eqnarray*}
and by choosing $K=2^{3n}$ we get that the statistical distance between $T_1$ and $T_{|\ip{\tau(L),w}| ,\lambda(L)^2}$
is exponentially small.

Recall that $w=\sum_{i=1}^n K^{i-1} v^*_i$ and $\tau(L)=\sum_{i=1}^n a_i v_i$ where all $|a_i| \le 2^{2n}$. Since
$\ip{v_i,v^*_j}=\delta_{ij}$, the inner product $\ip{\tau(L),w}$ is integer and its absolute value is at most $n\cdot
2^{2n}\cdot K^n \le 2^{\C{h} n^2}$ for a large enough $\C{h}$, as required.
\end{proof}

\section{Analysis of the Public Key Cryptosystem}\label{section_analysis_pke}

\begin{lemma}[Correctness]
The probability of a decryption error is at most $2^{-\Omega(\frac{(\gamma(n))^2}{m})}$ plus some exponentially small
terms.
\end{lemma}
Note that the above probability is negligible since $\gamma(n)=\omega(n \sqrt{\log n})$.
\begin{proof}
First consider an encryption of the bit $0$. Probabilities are taken over the choices of the private and public keys
and the randomization in the encryption process. Let $S$ denote the subset of indices which are included in the sum and
let $w:=\sum_{i\in S} a_i ~\mod~ N$. Since $\sum_{i\in S} a_i \le m\cdot N$,
$$\abs{w - (\sum_{i\in S} a_i ~\mod~ d \round{h})} \le m\cdot |N - d\round{h}|= m \cdot d \cdot \frc(h) < \frac{1}{16} d$$
and by the triangle inequality,
$$ \frc(\frac{w}{d}) <  \frac{1}{16} + \frc(\frac{\sum_{i\in S} a_i ~\mod~ d\round{h} }{d}) =
  \frac{1}{16} + \frc(\frac{\sum_{i\in S} a_i}{d}) <
  \frac{1}{16} + \frac{m}{d} + \frc(\frac{N}{d}\sum_{i\in S} z_i )$$
where the last inequality uses $|N\cdot z_i - a_i| < 1$. Notice that $\frc(\frac{N}{d}\sum_{i\in S} z_i ) =
\frc(\sum_{i\in S} (x_i+y_i) )= \frc(\sum_{i\in S} y_i )$. Hence,
$$ \frc(\frac{w}{d}) < \frac{1}{16}+\frac{m}{d} + \frc(\sum_{i\in S} y_i ) < \frac{1}{8}+ \frc(\sum_{i\in S} y_i )$$
where we used the fact that $d$ is much larger than $m$. With probability exponentially close to $1$, all $x_i$'s are
strictly less than $\ceil{h}-1$. Conditioned on that, the distribution of $y_i$ is $Q_\beta$ and the distribution of
$\sum_{i\in S}y_i~\mod~1$ is $Q_{|S|\beta}$ where $|S|\beta \le m\cdot \beta = O(\frac{m}{(\gamma(n))^2})$. Therefore,
according to Claim~\ref{tail_bound}, the probability of $\frc(\sum_{i\in S} y_i) > \frac{1}{16}$ is at most
$2^{-\Omega(\frac{\gamma(n)}{m})}$ and hence
 \begin{equation}\label{zero_enc}
 \frc(\frac{w}{d}) < \frac{1}{8}+\frac{1}{16}
 \end{equation}
which is less than $\frac{1}{4}$, as required.

The proof for the case of an encryption of $1$ is similar. By using the fact that $x_{i_0}$ is odd and that with
probability exponentially close to $1$, $\frc(y_{i_0})<\frac{1}{16}$ we get $\frc(\frac{\floor{a_{i_0}/2}}{d})
> \frac{1}{2} - \frac{1}{32} - \frac{1}{d}$. This, combined with (\ref{zero_enc}) gives
$$ \frc(\frac{w}{d}) > \frc(\frac{\floor{a_{i_0}/2}}{d}) - \frac{1}{8} - \frac{1}{16} > \frac{1}{4}$$
and the proof is completed.
\end{proof}

Before establishing the security of the construction, let us prove a few simple claims.

\begin{claim}\label{alternative_t}
For any $h\in \N, \beta\in \R$, let $X,Y$ be two independent random variables; $X$ is distributed uniformly over
$\{0,\frac{1}{h},\ldots,\frac{h-1}{h}\}$ and $Y$ is normal with mean 0 and variance $\frac{\beta}{2\pi h^2}$. Then
$T_{h,\beta}$ is equivalent to the distribution of the sum of $X$ and $Y$ reduced modulo $1$.
\end{claim}
\begin{proof}
\begin{eqnarray*}
 T_{h, \beta}(r) &=& Q_\beta(hr ~\mod~ 1) = \sum_{k=-\infty}^{\infty} \frac{1}{\sqrt{\beta}}
   e^{-\frac{\pi}{\beta}(hr-k)^2}=\\
 && \sum_{l=0}^{h-1} \sum_{k=-\infty}^{\infty} \frac{1}{\sqrt{\beta}} e^{-\frac{\pi}{\beta}(hr-hk-l)^2} =
 \sum_{l=0}^{h-1} \frac{1}{h} \sum_{k=-\infty}^{\infty} \frac{h}{\sqrt{\beta}} e^{-\frac{\pi h^2}{\beta}(r-k-\frac{l}{h})^2}
\end{eqnarray*}
\end{proof}

\begin{claim}\label{T_plus_N}
For $h\in \N$, $T_{h, \beta}+Q_\delta ~\mod~ 1 = T_{h,\beta+\delta h^2}$.
\end{claim}
\begin{proof}
According to Claim~\ref{alternative_t}, $T_{h,\beta}$ can be viewed as the sum of two random variables $X$ and $Y$
reduced modulo $1$. Therefore, $T_{h, \beta}+Q_\delta ~\mod~ 1 = X+Y+Q_\delta ~\mod~ 1$. But since both $Y$ and
$Q_\delta$ are normal, their sum modulo 1 is exactly $Q_{\frac{\beta}{h^2}+\delta}$ and we conclude the proof by using
Claim~\ref{alternative_t} again.
\end{proof}

\begin{definition}
Given a density function $X$ on $[0,1)$ we define its {\it compression} by a factor $\delta \ge 1$ as the distribution
on $[0,1)$ given by
$$ \frac{1}{\int_0^1 X(\delta x ~\mod~ 1) dx} X(\delta r ~\mod~ 1).$$
We denote the result by $C_\delta(X)$.
\end{definition}

Using the above definition, $T_{h,\beta}$ is a compression of $Q_\beta$ by a factor of $h$. Notice that if we can
sample efficiently from $X$ then we can also sample efficiently from its compression. This is done in a way similar to
that used to sample from $T_{h, \beta}$.

\begin{claim}\label{compression_T}
For any $h\in \N$ and $\delta\ge 1$, the compression of $T_{h, \beta}$ by a factor $\delta$ is $T_{\delta h, \beta}$.
\end{claim}

\begin{proof}
The proof follows directly from the definition of $T_{h,\beta}$.
\end{proof}
\begin{claim}\label{subset_sum_uniform}
For large enough $c$, when choosing $c\cdot l$ numbers $a_1,\ldots,a_{c\cdot l}$ uniformly from $0$ to $2^l-1$ the
probability that the statistical distance between the uniform distribution on $\{0,\ldots,2^l-1\}$ and the distribution
given by sums modulo $2^l$ of random subsets of $\{a_1,\ldots,a_{c\cdot l}\}$ is more than $2^{-l}$ is at most
$2^{-l}$.
\end{claim}
\begin{proof}
Let $X_{t,b}$ for $t\in \{0,\ldots,2^l-1\},b\in \{0,1\}^{c\cdot l}\setminus 0^{c\cdot l}$ denote the event that
$\sum_{i=1}^{c\cdot l} b_i a_i \equiv t ~(\mod~ 2^l)$ where the probability is taken over the choice of
$\{a_1,\ldots,a_{c\cdot l}\}$. Then, $E[X_{t,b}] = 2^{-l}$ and $V[X_{t,b}] < 2^{-l}$. Hence, $E[Y_t]=\frac{2^{c\cdot l}
- 1}{2^l} = 2^{(c-1)\cdot l} - 2^{-l}$ where $Y_t$ denotes $\sum_{b \in \{0,1\}^{c\cdot l}\setminus 0^{c\cdot l}}
X_{t,b}$. Moreover, for $b\neq b'$, the events $X_{t,b}$ and $X_{t,b'}$ are pairwise disjoint. Therefore, $V[Y_t] <
\frac{2^{c\cdot l} - 1}{2^l} < 2^{(c-1)\cdot l}$. Using the Chebyshev inequality,
$$ \Pr\left( \abs{Y_t - (2^{(c-1)\cdot l} - 2^{-l})} \ge 2^{(\frac{c-1}{2}+1)\cdot l} \right) \le 2^{-2l}$$
and hence,
$$ \Pr\left( \abs{Y_t - 2^{(c-1)\cdot l}} \ge 2^{(\frac{c-1}{2}+1)\cdot l} + 2^{-l} \right) \le 2^{-2l}.$$
 Using the union bound,
$$ \Pr\left( \exists t,~\abs{Y_t - 2^{(c-1)\cdot l}} \ge 2^{(\frac{c-1}{2}+1)\cdot l} + 2^{-l} \right) \le 2^{-l}.$$
Therefore, with probability at least $1 - 2^{-l}$ on the choice of $\{a_1,\ldots,a_{c\cdot l}\}$, the number of subsets
(including the empty subset) mapped to each number $t$ is at most $2^{(\frac{c-1}{2}+1)\cdot l}+2^{-l}+1 \le
2^{(\frac{c-1}{2}+2)\cdot l}$ away from $2^{(c-1)\cdot l}$. This translates to a statistical distance of at most
$$2^{(\frac{c-1}{2}+2)\cdot l} \cdot 2^{-(c-1)\cdot l} < 2^{-l}$$
for large enough $c$.
\end{proof}

\begin{lemma}[Security]
For $\C{N} \ge 2\C{h}$ and large enough $\C{m}$, if there exists a polynomial time algorithm $\calA$ that distinguishes
between encryptions of $0$ and $1$ then there exists an algorithm $\calB$ that distinguishes between the distributions
$U$ and $\calT_{n,\sqrt{n} \gamma(n)}$.
\end{lemma}
\begin{proof}
Let $p_0$ be the acceptance probability of $\calA$ on input $((a_1,\ldots,a_m,i_0),w)$ where $w$ is an encryption of
$0$ with the public key $(a_1,\ldots,a_m,i_0)$ and the probability is taken over the choice of private and public keys
and the encryption algorithm. We define $p_1$ similarly for encryptions of $1$ and let $p_u$ be the acceptance
probability of $\calA$ on inputs $((a_1,\ldots,a_m,i_0),w)$ where $a_1,\ldots,a_m,i_0$ are again chosen according to
the private and public keys distribution but $w$ is chosen uniformly from $\{0,\ldots,N-1\}$. We would like to
construct an $\calA'$ that distinguishes between the case where $w$ is an encryption of $0$ and the case where $w$ is
random. According to our hypothesis, $\abs{p_0-p_1} \ge \frac{1}{n^c}$ for some $c>0$. Therefore, either $\abs{p_0-p_u}
\ge \frac{1}{2n^c}$ or $\abs{p_1-p_u} \ge \frac{1}{2n^c}$. In the former case $\calA$ is itself the required
distinguisher. In the latter case $\calA$ distinguishes between the case where $w$ is an encryption of $1$ and the case
where $w$ is random. We construct $\calA'$ as follows. On input $((a_1,\ldots,a_n,i_0),w)$, $\calA'$ calls $\calA$ with
$((a_1,\ldots,a_n,i_0),w+\floor{\frac{a_{i_0}}{2}} ~\mod~ N)$. Notice that this maps the distribution on encryptions of
$0$ to the distribution on encryptions of $1$ and the uniform distribution to itself. Therefore, $\calA'$ is the
required distinguisher.

Let $p_0(a_1,\ldots,a_m,i_0)$ be the probability that $\calA'$ accepts on inputs $((a_1,\ldots,a_m,i_0),w)$ where the
probability is taken only over the choice of $w$ as an encryption of $0$ with the fixed public key
$(a_1,\ldots,a_m,i_0)$. Similarly, define $p_u(a_1,\ldots,a_m,i_0)$ to be the acceptance probability of $\calA'$ where
$w$ is now chosen uniformly at random from $\{0,\ldots,N-1\}$. Define
$$Y = \left\{ (a_1,\ldots,a_m,i_0) ~\left|~ |p_0(a_1,\ldots,a_m,i_0) - p_u(a_1,\ldots,a_m,i_0)| \ge \frac{1}{4n^c} \right. \right\}.$$
By an averaging argument we get that with probability at least $\frac{1}{4n^c}$ on the choice of
$(a_1,\ldots,a_m,i_0)$, $(a_1,\ldots,a_m,i_0)\in Y$ for otherwise $\calA'$ would have a gap of less than
$\frac{1}{2n^c}$.

In the following we describe the distinguisher $\calB$. We are given a distribution $R$ which is either $U$ or some
$T_{h,\beta}\in \calT_{n,\sqrt{n} \gamma(n)}$ with an integer $h\le 2^{\C{h} n^2} \le \sqrt{N}$ and a real $\beta\in
[\frac{1}{(\gamma(n))^2},4\frac{1}{(\gamma(n))^2})$. Note that neither $h$ nor $\beta$ are given to $\calB$. Our goal
is to construct $\calB$ such that the acceptance probability with $U$ and the acceptance probability with $T_{h,\beta}$
differ by a non-negligible factor. We first choose $\th$ uniformly from the set $\{1,2,4,\ldots,\sqrt{N}\}$. In
addition we choose $\delta$ uniformly from the range $[\sqrt{N}/\th, 4\sqrt{N}/\th)$ and $s$ uniformly from the range
$[0,7\frac{1}{(\gamma(n))^2})$. Then, consider the distribution $R' = C_\delta(R+Q_{\delta^2 s / N} ~\mod~ 1)$, i.e.,
we first add a normal variable to $R$ and then compress the result by a factor of $\delta$. We take $m$ samples
$a_1,\ldots,a_m$ from $\floor{N\cdot R'}$ and let $i_0$ be chosen randomly from $[m]$. We estimate
$p_0(a_1,\ldots,a_m,i_0)$ and $p_u(a_1,\ldots,a_m,i_0)$ by computing many values $w$ either according to the encryption
algorithm or randomly and then calling $\calA'$. By using a polynomial size sample, we can estimate the two
probabilities up to an error of at most $\frac{1}{32 n^c}$. If the two estimates differ by more than $\frac{1}{4 n^c}$,
$\calB$ accepts. Otherwise, $\calB$ rejects.

We first claim that when $R$ is the uniform distribution, $\calB$ rejects with high probability. The distribution
$R+Q_{\delta^2 s/N} ~\mod~ 1$ is still a uniform distribution on $[0,1)$ and so is $R'$ as can be easily seen from the
definition of the compression. Therefore, $a_1,\ldots, a_m$ are chosen uniformly from $\{0,1,\ldots, N-1\}$ and
according to Claim~\ref{subset_sum_uniform} if $\C{m}$ is a large enough constant then with probability exponentially
close to $1$, the distribution on $w$ obtained by encryptions of $0$ is exponentially close to the uniform distribution
on $\{0,1,\ldots,N-1\}$. Therefore, since $\calA'$ can be seen as a function on $w$,
$|p_0(a_1,\ldots,a_m,i_0)-p_u(a_1,\ldots,a_m,i_0)|$ is also exponentially small and $\calB$ rejects.

Now assume that $R$ is the distribution $T_{h,\beta}$ for some fixed $h$ and $\beta$ and we claim that $\calB$ accepts
with non-negligible probability. Then, according to Claim~\ref{T_plus_N}, $R+Q_{\delta^2 s/N} ~\mod~ 1$ is
$T_{h,\beta+(\delta h)^2 s/N}$. Hence, according to Claim~\ref{compression_T}, $R'$ is $T_{\delta h,\beta+(\delta
h)^2s/N}$. Let $X$ denote the event that $h\le \th< 2h$, $\delta h\in [\sqrt{N},2\sqrt{N})$, $\frc(\delta h) <
\frac{1}{16m}$ and $\beta+(\delta h)^2s/N \in [4\frac{1}{(\gamma(n))^2},8\frac{1}{(\gamma(n))^2})$. We now show that
the probability on our choice of $\th,\delta,s$ that $X$ happens is at least $\frac{1}{poly(n)}$. First, with
probability $\frac{1}{\log(\sqrt{N})}=\frac{2}{\C{N}n^2}$, $h\le \th< 2h$. Now, $\delta h$ is uniformly distributed in
$[h/\th \cdot \sqrt{N},4 h/\th \cdot \sqrt{N})$. Therefore, conditioned on $h\le \th< 2h$, the probability that $\delta
h \in [\sqrt{N},2\sqrt{N})$ is at least $\frac{1}{3}$. Moreover, conditioned on $h\le \th<2h$ and $\delta h \in
[\sqrt{N},2\sqrt{N})$, the probability that $\frc(\delta h)<\frac{1}{16m}$ is $\frac{1}{8m}$. For any fixed $\delta
h\in [\sqrt{N},2\sqrt{N})$, $(\delta h)^2/N \in [1,4)$ and therefore $\beta+(\delta h)^2 s/N$ is distributed uniformly
in $[\beta, \beta+(\delta h)^2 / N\cdot \frac{7}{(\gamma(n))^2})$. The length of this range is at most $4\cdot
\frac{7}{(\gamma(n))^2}$ and it always contains the range $[4\frac{1}{(\gamma(n))^2},8\frac{1}{(\gamma(n))^2})$
(because $\beta \in [\frac{1}{(\gamma(n))^2}, 4\frac{1}{(\gamma(n))^2})$). Therefore, the probability on the choice of
$s$ that $\beta+(\delta h)^2s/N \in [4\frac{1}{(\gamma(n))^2},8\frac{1}{(\gamma(n))^2})$ is at least
$\frac{4}{28}=\frac{1}{7}$. To sum up, the probability of $X$ is at least $\frac{2}{\C{N}n^2}\cdot \frac{1}{3}\cdot
\frac{1}{7} \cdot \frac{1}{8m} = \frac{1}{poly(n)}$.

Notice that conditioned on $X$, the distribution of $\delta h$ and $\beta+(\delta h)^2 /N$ is the same as the
distribution of $h$ and $\beta$ in the choice of the private and public keys. Therefore the probability that
$(a_1,\ldots,a_m,i_0) \in Y$ is at least
$$\Pr(X)\cdot \Pr \left( \exists i_0, (a_1,\ldots,a_m, i_0) \in Y ~|~ X \right) \cdot
   \frac{1}{m} \ge \Pr(X) \cdot \frac{1}{4n^c} \cdot \frac{1}{m} = \frac{1}{poly(n)}.$$
But when $(a_1,\ldots,a_m,i_0)\in Y$,
$$|p_0(a_1,\ldots,a_m,i_0)-p_u(a_1,\ldots,a_m,i_0)|\ge \frac{1}{4n^c}$$
and therefore our estimates are good enough and $\calB$ accepts.
\end{proof}

By combining the two lemmas above we get,

\begin{theorem}
For $\C{N} \ge 2\C{h}$ and large enough $\C{m}$, the public key cryptosystem described in
Section~\ref{section_crypto_constr} makes decryption errors with negligible probability and its security is based on
$\sqrt{n} \cdot \gamma(n)$-uSVP.
\end{theorem}

\section{Analysis of the Collision Resistant Hash Function}\label{section_analysis_hash}

\begin{claim}\label{normals_dot_vector}
Let $X_1,\ldots,X_m$ be $m$ independent normal random variables with mean 0 and standard deviation $\sigma$. For any
vector $b\in \R^m$, the random variable $\sum_{i=1}^m b_i X_i$ has a normal distribution with mean 0 and standard
deviation $\|b\|\cdot \sigma$.
\end{claim}
\begin{proof}
The joint distribution $(X_1,\ldots,X_m)$ is a Gaussian distribution in $\R^m$ which is invariant under rotations.
Hence we can equivalently consider the inner product of $(\norm{b},0,\ldots,0)$ and a Gaussian distribution. We
complete the proof by noting that the first coordinate of the Gaussian has a normal distribution with mean $0$ and
standard deviation $\sigma$.
\end{proof}

\begin{definition}
For any $h\in \Z$, $\th,\beta \in \R$ and any $a\in [0,1)$ we define the following two density functions on $[0,1)$:
$$ S_{\th,h,\beta,a}(r) := \frac{1}{\th\int_{a}^{a+1/\th} T_{h,\beta}(x) dx} T_{h,\beta}(a+\frac{r}{\th})$$
$$ S'_{h,\beta,a}(r) := T_{h,\beta}(a+\frac{r}{h}) = Q_\beta(a\cdot h + r~\mod~ 1).$$
\end{definition}

\begin{claim}\label{tilde_h_close_h}
If $h\le \th < (1+\delta)h$ where $h\in \Z$, $\th\in \R$, $\delta>0$ and $\beta \le \frac{1}{4}$ then
$\Delta(S_{\th,h,\beta,a},S'_{h,\beta,a}) \le \frac{\cc}{\beta}\delta$.
\end{claim}
\begin{proof}
According to Claim~\ref{sum_exp}, $T_{h,\beta}(x) = Q_\beta(h x ~\mod~ 1) \le (1+\sqrt{\beta})/\sqrt{\beta} \le
2/\sqrt{\beta}$ for any $x\in \R$. Therefore,
\begin{eqnarray*}
&& \int_a^{a+1/h} - \int_a^{a+1/\th} T_{h,\beta}(x)dx \le \frac{2}{\sqrt{\beta}}(\frac{1}{h} -
 \frac{1}{\th}) =
 \frac{2}{\sqrt{\beta}\cdot \th} (\frac{\th}{h} - 1) \le \frac{2\delta}{\sqrt{\beta}\cdot \th}
\end{eqnarray*}
But $\int_a^{a+1/h} T_{h,\beta}(x)dx = \frac{1}{h}$ and therefore we see that
$$ \frac{\th}{h} - \th \int_a^{a+1/\th}T_{h,\beta}(x)dx \le \frac{2\delta}{\sqrt{\beta}}.$$
Let $S''_{\th,h,\beta,a}(r) := T_{h,\beta}(a+r/\th)$. Then,
 \begin{eqnarray*}
 &&\int_0^1 \abs{S_{\th,h,\beta,a}(r) - S''_{\th,h,\beta,a}(r)} dr = \abs{1-\th \int_a^{a+1/\th}T_{h,\beta}(x)dx} \cdot \int_0^1 S_{\th,h,\beta,a}(r) dr = \\
 &&  \abs{1-\th \int_a^{a+1/\th}T_{h,\beta}(x)dx}
  \le \abs{1-\frac{\th}{h}} + \abs{\frac{\th}{h} - \th \int_a^{a+1/\th}T_{h,\beta}(x)dx} \le
 (1+\frac{2}{\sqrt{\beta}})\delta
 \end{eqnarray*}
Now, using the mean value theorem for any $r\in [0,1)$,
$$ \abs{S'_{h,\beta,a}(r) -  S''_{\th,h,\beta,a}(r)} \le
   (\frac{1}{h} - \frac{1}{\th}) \max_x \abs{\frac{d}{dx} T_{h,\beta}(x)} =
   (\frac{1}{h} - \frac{1}{\th}) \max_x \abs{\frac{d}{dx} \sum_{k=-\infty}^\infty
      \frac{1}{\sqrt{\beta}}e^{-\pi(\frac{h}{\sqrt{\beta}} x - \frac{1}{\sqrt{\beta}} k)^2}}
   $$
which, according to Claim~\ref{claim_normal_bound} using $\frac{1}{\sqrt{\beta}} \ge 2 > \frac{1}{\sqrt{2\pi}}+1$, is
at most
$$(\frac{1}{h}-\frac{1}{\th})\cdot \frac{\cc}{\sqrt{\beta}} \cdot \frac{h}{\sqrt{\beta}} = \frac{\cc}{\beta} (1-\frac{h}{\th}) \le \frac{\cc}{\beta}\delta.$$
To sum up,
\begin{eqnarray*}
 2\Delta(S_{\th,h,\beta,a},S'_{h,\beta,a}) &\le&
  \int_0^1 \abs{S_{\th,h,\beta,a}(r) - S''_{\th,h,\beta,a}(r)} dr + \int_0^1 \abs{S'_{h,\beta,a}(r) - S''_{\th,h,\beta,a}(r)}  dr\\
    &\le& (\frac{\cc}{\beta}+1+\frac{2}{\sqrt{\beta}}) \delta \le \frac{\cc}{\beta} \delta.
\end{eqnarray*}
\end{proof}

\begin{theorem}
For $\C{N} \ge 2\C{h}$ and any $\C{m} \ge 0$, if there exists an algorithm $\calA$ that given a list $a_1,\ldots,a_m
\in \{0,1,\ldots,N-1\}$ finds a nonzero vector $b \in \Z^m$ such that $\|b\| \le \sqrt{m}$ and $\sum_{i=1}^m b_i a_i
\equiv 0 (\mod ~ N)$ with probability at least $n^{-\C{a}}$ where $\C{a} > 0$ is some constant then there exists a
solution to $\sqrt{n} \cdot \gamma(n)$-uSVP.
\end{theorem}
Note that in particular, if $b \in \{-1,0,1\}^m$ then $\|b\| \le \sqrt{m}$ and hence this theorem includes collision
finding algorithms.
\begin{proof}
According to Theorem~\ref{thm_main_theorem} it is enough to construct a distinguisher $\calB$ between $U$ and
$\calT_{n,\sqrt{n}\cdot \gamma(n)}$. The distinguisher $\calB$ works by calling the routine $\calC$ described below $n$
times with each value $\th = (1+n^{-\C{\th}})^i$, $i\in [\log_{1+n^{-\C{\th}}} N]$. The constant $\C{\th}$ will be
specified later. If there exists an $\th$ for which all $n$ calls to $\calC$ accept, $\calB$ accepts. Otherwise, for
any $\th$ there exists one call where $\calC$ rejects and $\calB$ rejects.

The routine $\calC(\th)$ samples $m$ values $x_1,\ldots,x_m$ from the given distribution which we denote by $R$. It
also chooses $m$ values $y_1,\ldots,y_m$ uniformly in $[0,1/\th)$. Let $z_i=x_i-y_i ~\mod~ 1$ and $a_i = \floor{N\cdot
z_i}$. We call $\calA$ with $a_1, \ldots, a_m$. If $A$ fails we repeat the process again (choosing $x_i, y_i$ and
calling $\calA$). If after $n^{\C{a}+1}$ calls $\calA$ still fails, $\calC$ accepts. Otherwise, we have a vector $b\in
\Z^m$ such that $\|b\| \le \sqrt{m}$ and $\sum_{i=1}^m b_i a_i \equiv 0 (\mod~ N)$. The routine $\calC(\th)$ accepts if
$\frc(\sum_{i=1}^m b_i \th y_i) < \frac{1}{4}$ and rejects otherwise.

First we show that if $R$ is the uniform distribution then for any $\th$, $\calC(\th)$ accepts with probability roughly
$\frac{1}{2}$. From this it will follow that the probability that $n$ calls to $\calC(\th)$ accept is exponentially
small, i.e., $\calB$ rejects with probability exponentially close to $1$. Each number $x_i$ is uniform in $[0,1)$ and
so is $z_i$. Therefore, each $a_i$ is uniform in $\{0,1,\ldots,N-1\}$ and according to our assumption, $\calA$ succeeds
with probability at least $n^{-\C{a}}$. The probability that $n^{\C{a}+1}$ calls fail is at most
$(1-n^{-\C{a}})^{n^{\C{a}+1}} < e^{-n}$ which is exponentially small. In order to bound the probability that
$\frc(\sum_{i=1}^m b_i \th y_i) < \frac{1}{4}$ we use the fact that $\calA$ is oblivious to the decomposition of the
$z_i$'s into $x_i-y_i$ and would work equally well if $z_i=x_i'-y_i'$ for some other $x_i'$,$y_i'$. Consider the
following equivalent way to create the joint distribution of $x_i, y_i, z_i$: we first choose the $z_i$'s uniformly in
$[0,1)$ and then choose $y_i$ uniformly in $[0,1/\th)$ and choose $x_i$ to be $z_i + y_i ~\mod~ 1$. Hence, conditioned
on any values for the $z_i$'s, the distribution of the $y_i$'s is uniform in $[0,1/\th)$ and therefore
$\frc(\sum_{i=1}^m b_i \th y_i)$ is distributed uniformly in $[0,\frac{1}{2})$. The probability that $\frc(\sum_{i=1}^m
b_i \th y_i) < \frac{1}{4}$ is therefore $\frac{1}{2}$, as required.

Now consider the case that $R$ is $T_{h,\beta}$ where $\beta \le \frac{4}{(\gamma(n))^2}$. We claim that when $\th$ is
the smallest such that $\th \ge h$, $\calC(\th)$ rejects with probability at most $\cc m n^{2\C{\gamma} - \C{\th}}$.
Therefore, the probability that $\calB$ sees a rejection after $n$ calls is at most $\cc m n^{2\C{\gamma} - \C{\th}+1}$
and it therefore accepts with probability close to $1$ if we choose a large enough $\C{\th}$. Notice that such an $\th$
satisfies $h\le \th < (1+n^{-\C{\th}}) h$. As before, we create the joint distribution of $x_i,y_i,z_i$ by first
choosing $z_i$ and then $y_i$. This would allow us to use the fact that $\calA$ is oblivious to the decomposition of
$z_i$ to $x_i-y_i$. So we first choose the $z_i$'s from their unconditional distribution and then consider the
distribution of $y_i$ conditioned on $z_i$ given by:
$$ \frac{1}{\int_{z_i}^{z_i+1/\th} T_{h,\beta}(x) dx} T_{h,\beta}(z_i+r) \quad \forall r\in [0,\frac{1}{\th}).$$
Hence the density function of the distribution of $\th\cdot y_i$ is exactly $S_{\th,h,\beta,z_i}$. According to
Claim~\ref{tilde_h_close_h} the statistical distance between $S_{\th,h,\beta,z_i}$ and $S'_{h,\beta,z_i}$ is at most
$\frac{\cc}{\beta}n^{-\C{\th}} \le \cc n^{2\C{\gamma}-\C{\th}}$. Let $\xi_1,\ldots,\xi_m$ be $m$ random variables
chosen independently according to $Q_\beta$. Notice that the distribution of the random variable $\xi_i - h\cdot z_i$
is exactly $S'_{h,\beta,z_i}$. Hence, according to Claim~\ref{joint_statistical_distance} the statistical distance
between the joint distributions $(\th \cdot y_1,\ldots,\th \cdot y_m)$ and $(\xi_1 - h\cdot z_1, \ldots, \xi_m - h\cdot
z_m)$ is at most $\cc m\cdot n^{2\C{\gamma}-\C{\th}}$. Now,
$$ \sum_{i=1}^m b_i (\xi_i - h\cdot z_i) ~\mod~ 1 = \sum_{i=1}^m b_i \xi_i - \sum_{i=1}^m b_i \cdot h \cdot z_i ~\mod~ 1 $$
According to Claim~\ref{normals_dot_vector}, $\sum_{i=1}^m b_i \xi_i$ has a normal distribution with mean 0 and
standard deviation $\|b\| \cdot \sqrt{\frac{\beta}{2\pi}} \le \sqrt{\frac{m\beta}{2\pi}} \le \sqrt{\frac{2m}{\pi
(\gamma(n))^2}} = o(\frac{1}{\sqrt{\log n}}).$ Therefore, according to Claim~\ref{tail_bound}, the probability that
$\frc(\sum_{i=1}^m b_i \xi_i) > \frac{1}{8}$ is negligible. Now,
$$\frc(\sum_{i=1}^m b_i \cdot h \cdot z_i) =
\frc(\sum_{i=1}^m \frac{h}{N} b_i N\cdot z_i ) \le \frc(\sum_{i=1}^m \frac{h}{N} b_i a_i ) +
\sum_{i=1}^m\frac{h}{N}\cdot b_i = \sum_{i=1}^m\frac{h}{N}\cdot b_i \le m\cdot \frac{h}{N} \cdot \sqrt{m}.$$
 Therefore, except with negligible probability,
 $$ \frc(\sum_{i=1}^m b_i (\xi_i - h\cdot z_i)) \le \frac{1}{8} + m\cdot \frac{h}{N} \sqrt{m} < \frac{1}{4}$$
where we used the fact that $h\le 2^{\C{h} n^2} \le \sqrt{N}$. This implies that the probability that
$\frc(\sum_{i=1}^m b_i \th y_i)< \frac{1}{4}$ is at most $\cc m\cdot n^{2\C{\gamma}-\C{\th}}$ plus some negligible
amount.
\end{proof}

\section{Quantum Computation}\label{section_quantum}

In this section we show a result related to a problem in quantum computation known as the dihedral hidden subgroup
problem. One reason this problem is interesting is because, under certain conditions, solving it implies a quantum
solution to uSVP~\cite{RegevQuantumLattices}. In~\cite{EttingerHoyer}, Ettinger and H{\o}yer reduced the problem to the
problem of finding an integer $k$ given access to the distribution $Z_k$ where $\Pr(Z_k=z) = 2/N \cdot \cos^2 (\pi k z
/ N)$ for $z=0,1,\ldots,N-1$. They presented an exponential time classical algorithm that uses only a polynomial number
of samples of $Z_k$. Hence, a polynomial number of samples contains enough information to find $k$. The question of
whether there exists an {\em efficient} algorithm remained open. In this section we will show that a solution to their
problem implies a solution to $n^c$-uSVP for some $c$.

We start by extending Theorem \ref{thm_main_theorem} to more general periodic distributions. Let $D$ be a distribution
on $[0,1)$ such that its density function satisfies $D(r) \le \C{D}$ and $|D(r)-D(r+\epsilon ~\mod~ 1)| \le \C{D}
\epsilon$ for all $r,\epsilon \in [0,1)$ for some constant $\C{D}$. For $h\in \N$, define
$$ T^D_h(r)=D(rh ~\mod~ 1)$$
to be the distribution on $[0,1)$ given by $h$ periods of $D$. Moreover, define
$$ \calT^D_n = \left\{ T^D_h ~|~ h\in \N,~ h\le 2^{\C{h}n^2} \right\}.$$
where $\C{h}$ is the constant from Lemma~\ref{lemma_one_dim_distr} and $n$ is the size parameter of the problem.

\begin{lemma}\label{lm:quantum}
If there exists a distinguisher between $U$ and $\calT^D_n$ then there exists a solution to $n^c$-uSVP for some $c>0$.
\end{lemma}
\begin{proof}
Assume $\calA$ is a distinguisher between $U$ and $\calT^D_n$ and assume that it uses $n^{\C{A}}$ samples of the given
distribution for some $\C{A}>0$. Let $p_u$ denote the acceptance probability of $\calA$ on inputs from distribution $U$
and for $h\in \N$ let $p_h$ denote its acceptance probability on inputs from $T^D_h$. According to our hypothesis $|p_u
- p_h| \ge n^{-\C{d}}$ for all $h\in [2^{\C{h}n^2}]$ for some constant $\C{d}>0$.

We construct a distinguisher $\calB$ between $U$ and $\calT_{n,n^c}$ for some large enough $c>0$. The lemma then
follows from Theorem \ref{thm_main_theorem}. Let $R$ denote the given distribution. First, $\calB$ chooses a value
$\th$ uniformly from the set $\{1,1+\mu,(1+\mu)^2,\ldots,2^{\C{h}n^2}\}$ where $\mu=n^{-\C{\mu}}$ for some constant
$\C{\mu}>0$ to be chosen later. Then, define the distribution $R'$ as
$$ R'=R+\frac{D}{\th} ~\mod~ 1,$$
i.e., a sample from $R'$ is given by $x+r/\th ~\mod~ 1$ where $x$ is chosen from $R$ and $r$ is chosen from $D$. It
then estimates the acceptance probability of $\calA$ using sequences of samples from $R'$ each of length $n^{\C{A}}$.
According to the Chernoff bound, using a polynomial number of sequences, we can obtain an estimate that with
probability exponentially close to 1 is within $\frac{1}{4n^{\C{d}}}$ of the actual acceptance probability. If the
estimate differs from $p_u$ by more than $\frac{1}{2n^{\C{d}}}$, $\calB$ accepts; otherwise, it rejects. This completes
the description of $\calB$.

When $R$ is the uniform distribution then $R'$ is also the uniform distribution. Therefore, with probability
exponentially close to $1$, $\calB$'s estimate is within $\frac{1}{4n^{\C{d}}}$ of $p_u$ and $\calB$ rejects. Hence, it
is remains to show that $\calB$ accepts with some non-negligible probability when $R$ is $T_{h,\beta}$ where $h\le
2^{\C{h}n^2}$ and $\beta \le n^{-\C{\beta}}$ for some large enough $\C{\beta}$.

Consider the event in which $h\le \th < (1+\mu)h$. Notice that it happens with non-negligible probability since $\th$
is chosen from a set of size polynomial in $n$. The following claim will complete the proof by showing that the
statistical distance between $R'$ and $T^D_h$ is smaller than $n^{-\C{A}-\C{d}}/4$. Using Claim
\ref{joint_statistical_distance}, it follows that the statistical distance between a sequence of $n^{\C{A}}$ elements
of $R'$ and a sequence of $n^{\C{A}}$ elements of $T^D_h$ is at most $n^{-\C{d}}/4$. Finally, using Equation
\ref{eq:statistical_dst}, this implies that $\calA$'s success probability on sequences from $R'$ is within
$n^{-\C{d}}/4$ from $p_h$ and since $|p_u-p_h| \ge n^{-\C{d}}$, $\calB$ accepts.

\begin{claim}
For $\th$ as above and for large enough $\C{\beta}$ and $\C{\mu}$, the statistical distance $\Delta(R',T^D_h) \le
n^{-\C{A}-\C{d}}/4$.
\end{claim}
\begin{proof}
Consider the distribution $R''$ given by
$$ R'' = T_{h,\beta}+\frac{D}{h}.$$
The distribution $R''$ can be seen as a random function of the distribution $D$: given a value $r\in D$ sample a value
$x$ from $T_{h,\beta}$ and output $x+r/h$. Notice that $R'$ is given by applying the same function to the distribution
$(h/\th) D$. Hence, using Equation \ref{eq:statistical_dst},
\begin{eqnarray}
 \Delta(R',R'') \le \Delta(D, \frac{h}{\th} D) &=& \int_0^{h/\th} |D(r)-D(\frac{\th}{h}r)| dr + \int_{h/\th}^1 D(r)dr \nonumber \\
 &\le& \C{D} (1-\frac{h}{\th}) + (1-\frac{h}{\th})\C{D} \nonumber \\
 &\le& 2\C{D} \mu = 2\C{D} n^{-\C{\mu}}. \label{eq:quantum1}
\end{eqnarray}

We next bound the statistical distance between $T^D_h$ and $R''$. Let $X$ be a random variables distributed uniformly
over $\{0,\frac{1}{h},\ldots,\frac{h-1}{h}\}$. Then, it can be seen that
$$T^D_h = X + \frac{D}{h} ~\mod~ 1.$$
Now, let $Y$ be another random variable distributed normally with mean 0 and variance $\frac{\beta}{2\pi}$. Then, as in
Claim \ref{alternative_t}, $T_{h,\beta}=X+Y/h ~\mod~ 1$ and hence,
\begin{eqnarray*}
R'' &=& X + \frac{Y}{h} + \frac{D}{h} ~\mod~ 1.
\end{eqnarray*}
Therefore, $T^D_h$ can be seen as a random function applied to a sample from $\frac{D}{h}$ while $R''$ can be seen as
the same function applied to a sample from $\frac{Y}{h} + \frac{D}{h}$. From Equation \ref{eq:statistical_dst} it
follows that
\begin{equation}\label{eq:quantum2}
\Delta(T^D_h,R'') \le \Delta \left( \frac{1}{h}D ,~ \frac{1}{h}(D+Y) \right) = \Delta(D,D+Y).
\end{equation}
Let $Y'$ be the restriction of a normal distribution with mean 0 and variance $\frac{\beta}{2\pi}$ to the range
$[-n\sqrt{\beta}, n\sqrt{\beta}]$. More formally,
$$ Y'(r) = Y(r) / \int_{-n\sqrt{\beta}}^{n\sqrt{\beta}} Y(r) dr $$
for $r \in [-n\sqrt{\beta}, n\sqrt{\beta}]$ and $Y'(r)=0$ elsewhere. From Claim \ref{tail_bound} it follows that the
distribution of $Y$ is very close to that of $Y'$:
\begin{equation}\label{eq:quantum3}
\Delta(Y, Y') \le \sqrt{\frac{2}{\pi}} \cdot \frac{1}{n\sqrt{2\pi}} \cdot e^{- \pi n^2} = 2^{-\Omega(n^2)}.
\end{equation}
Now, using the fact that $Y'$ always gets values of small absolute value,
\begin{eqnarray*}
\left| D(r) - (D+Y')(r) \right| &=& \left| D(r) - \int_{-n\sqrt{\beta}}^{n\sqrt{\beta}}D(r-x)Y'(x)dx \right| \\
&\le& \int_{-n\sqrt{\beta}}^{n\sqrt{\beta}} \left| D(r)-D(r-x) \right| Y'(x)dx  \\
&\le& \C{D}n\sqrt{\beta} \int_{-n\sqrt{\beta}}^{n\sqrt{\beta}} Y'(x)dx \\
&=& \C{D}n\sqrt{\beta}.
\end{eqnarray*}
Since both $D(r)$ and $(D+Y')(r)$ are zero for $r < -n\sqrt{\beta}$ and for $r > 1+n\sqrt{\beta}$,
\begin{eqnarray}
\Delta(D,D+Y') &=& \int_{-n\sqrt{\beta}}^{1+n\sqrt{\beta}} \left| D(r) - (D+Y')(r) \right| dr \nonumber \\
 &\le& (1+2n\sqrt{\beta}) \cdot \C{D}n\sqrt{\beta} \nonumber \\
 &\le& (1+2 n^{1-\C{\beta}/2}) \cdot \C{D}n^{1-\C{\beta}/2} \le 2 \C{D}n^{1-\C{\beta}/2} \label{eq:quantum4}
\end{eqnarray}
for large enough $\C{\beta}$. Finally, combining Equations \ref{eq:quantum1}, \ref{eq:quantum2}, \ref{eq:quantum3},
\ref{eq:quantum4} and using the triangle inequality, we obtain
$$ \Delta(R', T^D_h) \le 2\C{D} n^{-\C{\mu}} + 2^{-\Omega(n^2)} + 2 \C{D}n^{1-\C{\beta}/2} \le n^{-\C{A}-\C{d}}/4$$
for large enough $\C{\beta}$ and $\C{\mu}$.
\end{proof}
This completes the proof of Lemma \ref{lm:quantum}.
\end{proof}

We can now prove the main theorem of this section.

\begin{theorem}
For $k \in \N$, $k < N$, define the distribution $Z_k$ by $\Pr(Z_k=z) = 2/N \cdot \cos^2 (\pi k z / N)$ for
$z=0,1,\ldots,N-1$. Assume there exists an algorithm $\calA$ that given a polynomial (in $\log N$) number of samples
from $Z_k$, returns $k$ with probability exponentially close to 1. Then, there exists a solution to $n^c$-uSVP for some
$c$.
\end{theorem}
We remark that it is possible to relax the assumptions of the theorem. It is enough if the algorithm returns $k$ with
non-negligible probability. Also, it is enough if the algorithm finds $k$ only for some non-negligible fraction of all
possible $k$'s.
\begin{proof}
Let $D$ be the distribution on $[0,1)$ given by $D(r)=2 \cos^2 (\pi r)$. An easy calculation shows that the absolute
value of its derivative is at most $4\pi$. Therefore, it satisfies the conditions stated before Lemma \ref{lm:quantum}
with $\C{D}=4\pi$. Using Lemma \ref{lm:quantum}, it is enough to show how to distinguish between $U$ and $\calT^D_n$.

Given an unknown distribution $R$, let $R'$ be the distribution given by $\lfloor N\cdot R\rfloor$ where $N$ is chosen
to be large enough, say, $2^{2\C{h}n^2}$. We call $\calA$ with enough samples from $R'$ and obtain a value $k$.
Finally, we take one sample $r$ from $R$ and accept if $\frc(rk) < 1/4$ and reject otherwise.

First, consider the case where $R$ is the uniform distribution. Then no matter which value of $k$ we obtain, the
probability that $\frc(rk) < 1/4$ is exactly $1/2$. Now consider the case where $R$ is $T^D_h$ for some $h \le
2^{\C{h}n^2}$. For any $r=0,\ldots,N-1$, the probability that $R'=r$ is given by
$$ \int_{r/N}^{(r+1)/N} D(hx ~\mod~ 1) dx =  \int_{r/N}^{(r+1)/N} 2 \cos^2 (\pi h x) dx.$$
From the bound on the derivative of $D$ mentioned above, we obtain that the distance of this integral from $2/N \cdot
\cos^2 (\pi h r / N)$ is at most $4\pi^2 h/N^2$. Therefore, the statistical distance between $R'$ and $Z_h$ is
$$ \Delta(Z_h, R') \le \frac{N}{2} \cdot 4\pi^2 h/N^2 = 2^{-\Omega(n^2)}.$$
Since the number of samples given to $\calA$ is only polynomial in $n$, its input is still within statistical distance
$2^{-\Omega(n^2)}$ of $Z_h$ and it therefore outputs $h$ with probability exponentially close to 1. Then, the
probability that $\frc(rk) < 1/4$ is given by
$$ \int_{-1/4}^{1/4} 2 \cos^2(\pi r) dr = \frac{1}{2} + \frac{1}{\pi}.$$
\end{proof}

\section{Acknowledgements}

I thank Irit Dinur for suggesting that I look at cryptographic constructions and Daniele Micciancio for many helpful
comments on an earlier draft of this paper.

\bibliographystyle{plain}
\bibliography{harmonic}

\begin{thebibliography}{10}

\bibitem{Ajtai96}
M.~Ajtai.
\newblock Generating hard instances of lattice problems.
\newblock In {\em Proc. 28th ACM Symp. on Theory of Computing}, pages 99--108,
  1996.

\bibitem{AjtaiDwork}
M.~Ajtai and C.~Dwork.
\newblock A public-key cryptosystem with worst-case/average-case equivalence.
\newblock In {\em Proc. 29th ACM Symp. on Theory of Computing}, pages 284--293,
  1997.

\bibitem{Banaszczyk}
W.~Banaszczyk.
\newblock New bounds in some transference theorems in the geometry of numbers.
\newblock {\em Mathematische Annalen}, 296(4):625--635, 1993.

\bibitem{CaiAW}
J.-Y. Cai.
\newblock Applications of a new transference theorem to {A}jtai's connection
  factor.
\newblock In {\em Fourteenth Annual IEEE Conference on Computational Complexity
  (Atlanta, GA, 1999)}, pages 205--214. IEEE Computer Soc., Los Alamitos, CA,
  1999.

\bibitem{CaiNerurkarAW}
J.-Y. Cai and A.~P. Nerurkar.
\newblock An improved worst-case to average-case connection for lattice
  problems (extended abstract).
\newblock In {\em 38th Annual Symposium on Foundations of Computer Science},
  pages 468--477, 1997.

\bibitem{EttingerHoyer}
M.~Ettinger and P.~H{\o}yer.
\newblock On quantum algorithms for noncommutative hidden subgroups.
\newblock {\em Advances in Applied Mathematics}, 25(3):239--251, 2000.

\bibitem{GolGolHal96}
O.~Goldreich, S.~Goldwasser, and S.~Halevi.
\newblock Collision-free hashing from lattice problems.
\newblock In {\em ECCCTR: Electronic Colloquium on Computational Complexity,
  technical reports}, 1996.

\bibitem{GoldreichGH97}
O.~Goldreich, S.~Goldwasser, and S.~Halevi.
\newblock Eliminating decryption errors in the {Ajtai-Dwork} cryptosystem.
\newblock {\em Lecture Notes in Computer Science}, 1294:105, 1997.

\bibitem{GoldreichGH97b}
O.~Goldreich, S.~Goldwasser, and S.~Halevi.
\newblock Public-key cryptosystems from lattice reduction problems.
\newblock In {\em Advances in cryptology (CRYPTO) '97 (Santa Barbara, CA,
  1997)}, volume 1294 of {\em Lecture Notes in Comput. Sci.}, pages 112--131.
  Springer, 1997.

\bibitem{NTRU}
J.~Hoffstein, J.~Pipher, and J.~H. Silverman.
\newblock N{TRU}: a ring-based public key cryptosystem.
\newblock In {\em Algorithmic number theory (Portland, OR, 1998)}, volume 1423
  of {\em Lecture Notes in Comput. Sci.}, pages 267--288. Springer, 1998.

\bibitem{ImpagliazzoNaor}
R.~Impagliazzo and M.~Naor.
\newblock Efficient cryptographic schemes provably as secure as subset sum.
\newblock {\em J. Cryptology}, 9(4):199--216, 1996.

\bibitem{LLL}
A.~K. Lenstra, H.~W. Lenstra, Jr., and L.~Lov{\'a}sz.
\newblock Factoring polynomials with rational coefficients.
\newblock {\em Math. Ann.}, 261(4):515--534, 1982.

\bibitem{Micciancio01hnf}
D.~Micciancio.
\newblock Improving lattice based cryptosystems using the hermite normal form.
\newblock In {\em Cryptography and Lattices Conference (CaLC)}, volume 2146 of
  {\em Lecture Notes in Computer Science}, pages 126--145, Providence, Rhode
  Island, March 2001. Springer-Verlag.

\bibitem{Mic02cyclic}
D.~Micciancio.
\newblock Generalized compact knapsacks, cyclic lattices, and efficient one-way
  functions from worst-case complexity assumptions.
\newblock In {\em Proceedings of the 43rd Annual Symposium on Foundations of
  Computer Science (FOCS) 2002}, Vancouver, Canada, November 2002.

\bibitem{Micciancio02hash}
D.~Micciancio.
\newblock Improved cryptographic hash functions with worst-case/average-case
  connection.
\newblock In {\em Proceedings of the 34th Annual ACM Symposium on Theory of
  Computing (STOC) 2002}, pages 609--618, Montreal, Canada, 2002.

\bibitem{MicciancioBook}
D.~Micciancio and S.~Goldwasser.
\newblock {\em Complexity of Lattice Problems: a cryptographic perspective},
  volume 671 of {\em The Kluwer International Series in Engineering and
  Computer Science}.
\newblock Kluwer Academic Publishers, Boston, Massachusetts, March 2002.

\bibitem{RegevQuantumLattices}
O.~Regev.
\newblock Quantum computation and lattice problems.
\newblock In {\em Proceedings of the 43rd Annual Symposium on Foundations of
  Computer Science (FOCS) 2002}, Vancouver, Canada, November 2002.

\bibitem{StefankovicThesis}
D.~\v{S}tefankovi\v{c}.
\newblock Fourier transforms in computer science.
\newblock Master's Thesis, University of Chicago, Department of Computer
  Science, TR-2002-03.

\end{thebibliography}

\appendix

\section{Several Technical Claims}\label{appendix_technical_claims}

\begin{claim}\label{tail_bound}
The probability that the distance of a normal variable with variance $\sigma^2$ from its mean is more than $t$ is at
most $\sqrt{\frac{2}{\pi}}\cdot \frac{\sigma}{t} e^{-\frac{t^2}{2\sigma^2}}$.
\end{claim}
\begin{proof}
$$\int_t^\infty \frac{1}{\sqrt{2\pi} \sigma} e^{-\frac{x^2}{2\sigma^2}} dx \le
  \int_t^\infty (1+\frac{\sigma^2}{x^2})\frac{1}{\sqrt{2\pi} \sigma} e^{-\frac{x^2}{2\sigma^2}} dx =
  -\frac{1}{\sqrt{2\pi}\sigma} \cdot \frac{\sigma^2}{x} e^{-\frac{x^2}{2\sigma^2}}\bigg|_{x=t}^\infty =
  \frac{\sigma}{\sqrt{2\pi}t} e^{-\frac{t^2}{2\sigma^2}}.$$
\end{proof}

\begin{claim}\label{sum_exp}
$$ \forall x,r\in \R, ~~ \sum_{k \in \Z} e^{-\pi (kr+x)^2} \le 1 + \frac{1}{r} $$
\end{claim}
\begin{proof}
Let $k'\in \Z$ be such that $|kr+x|$ is minimized. Then,
\begin{eqnarray*}
 && \sum_{k \in \Z} e^{-\pi (kr+x)^2} \le 1+ \sum_{k \in \Z \setminus \{k'\}} e^{-\pi (kr+x)^2} =
    1+ \frac{1}{r}\sum_{k \in \Z \setminus \{k'\}} r \cdot e^{-\pi (kr+x)^2} \le \\
 && 1+ \frac{1}{r}\int_{-\infty}^\infty e^{-\pi y^2} dy = 1+\frac{1}{r}
\end{eqnarray*}
where changing the sum to an integral is possible because the sum can be seen as the area under a function that lies
completely below $e^{-\pi y^2}$.
\end{proof}

\begin{claim}\label{claim_normal_bound}
For any $a,x\in \R$ and any $b > \frac{1}{\sqrt{2\pi}}+1$, $\abs{\frac{d}{dx} \sum_{k\in \Z} e^{-\pi (b k+a x)^2}} \le
\cc a $
\end{claim}
\begin{proof}
Let $z$ denote $a\cdot x$. Then,
\begin{eqnarray*}
&& \abs{\frac{d}{dx} \sum_{k\in \Z} e^{-\pi (bk+ax)^2}} = a\abs{\frac{d}{dz} \sum_{k\in \Z} e^{-\pi (bk+z)^2}} = \\
&& a\abs{\sum_{k\in \Z} -2\pi (bk+z) e^{-\pi (bk+z)^2}} \le \\
&& a\sum_{k\in \Z} \abs{2\pi (bk+z) e^{-\pi (bk+z)^2}}
\end{eqnarray*}
In the following we will upper bound
$$\sum_{k\in \{0,1,\ldots\}} \abs{2\pi (bk+y) e^{-\pi (bk+y)^2}}$$
where $y\in [0,b]$. The upper bound for the original expression is clearly at most $2a$ times this value. The function
$|2\pi y e^{-\pi y^2}|$ is increasing from $0$ to $\frac{1}{\sqrt{2\pi}}$ where it attains the maximum value of
$\sqrt{2\pi e}$. After that point it is monotonically decreasing. Hence,
\begin{eqnarray*}
&& \sum_{k\in \N} \abs{2\pi (bk+y) e^{-\pi (bk+y)^2}} = \\
&& \abs{2\pi y e^{-\pi y^2}} + \sum_{k\in \{1,2,\ldots\}} \abs{2\pi (bk+y) e^{-\pi (bk+y)^2}} \le \\
&& \sqrt{2\pi e} + \int_0^\infty \abs{2\pi y e^{-\pi y^2}} dy = \sqrt{2\pi e} + 1
\end{eqnarray*}
where changing from summation to integration is possible because $b\ge 1$ and because the function is decreasing from
$\frac{1}{\sqrt{2\pi}}$ and the first $y$ in the sum is at least $\frac{1}{\sqrt{2\pi}}+1$.
\end{proof}

\begin{claim}\label{cl:integral_periodic}
Let $L$ be a lattice and let $f:\R^n \rightarrow \R$ be periodic on $L$, i.e., $f(x)=f(x+y)$ for all $x\in \R^n$ and
$y\in L$. Then, for any two bases $v_1,\ldots,v_n$ and $u_1,\ldots,u_n$ of $L$,
$$ \int_{\calP(v_1,\ldots,v_n)} f(x) dx =  \int_{\calP(u_1,\ldots,u_n)} f(x) dx.$$
\end{claim}
\begin{proof}
One can get from one basis of a lattice to any other by a finite sequence of operations of the following two types:
replace vector $v_i$ by $-v_i$ and replace vector $v_i$ by $v_i+v_j$ for some $i\neq j$. Hence, it is enough to show
that the integral is invariant under these two operations. Define the following `half' parallelepipeds:
$$ \calP_1 = \{ \sum_{i=1}^n \alpha_i v_i ~|~ \alpha_i \in [0,1), ~ \alpha_2 \ge \alpha_1 \} $$
$$ \calP_2 = \{ \sum_{i=1}^n \alpha_i v_i ~|~ \alpha_i \in [0,1), ~ \alpha_2 < \alpha_1 \} $$
$$ \calP_3 = \{ \sum_{i=1}^n \alpha_i v_i +v_2 ~|~ \alpha_i \in [0,1), ~ \alpha_2 < \alpha_1 \} $$
Note that $\calP(v_1,\ldots,v_n)$ is equal to $\calP_1 \cup \calP_2$ and $\calP(v_1+v_2,v_2,\ldots,v_n) = \calP_1 \cup
\calP_3$. But since $\calP_3$ is a shift of $\calP_2$ by $v_2\in L$,
 $$ \int_{\calP(L)}f(x) dx = \int_{\calP_1}+\int_{\calP_2} f(x)dx = \int_{\calP_1}+\int_{\calP_3} f(x)dx =
 \int_{\calP(v_1+v_2,v_2,\ldots,v_n)} f(x)dx.$$
A similar argument shows that the integral is invariant under negation of basis vectors.
\end{proof}

\begin{claim}\label{t_density_function}
For any vector $v\in L$,
 $$\int_{\calP(L^*)} \sum_{k\in \Z} e^{-\pi (\frac{k + \ip{v,x}}{\|v\|})^2} dx = \|v\| d(L^*).$$
\end{claim}
\begin{proof}
Define
 $$f(x) := \sum_{k\in \Z} e^{-\pi (\frac{k + \ip{v,x}}{\|v\|})^2}.$$
Notice that for any $w\in L^*$, $f(x)=f(x+w)$ since $\ip{v,w} \in \Z$ and hence $f$ is periodic on $L^*$. Consider the
basis for $L^*$ given by any basis of the lattice $L^*\cap v^\bot$ and any vector $w$ in $L^*$ such that $\ip{w,v} =
1$. Let $\calP$ denote the corresponding parallelepiped. Then, using Claim \ref{cl:integral_periodic},
\begin{eqnarray*}
 \int_{\calP(L^*)} f(x) dx &=& \int_{\calP} f(x) dx =
 \frac{1}{\|v\|} \int_0^1 \int_{\calP \cap \{y| \ip{y,v} = a\}} f(x) dx~da \\
 &=& \frac{1}{\|v\|} \int_0^1 \int_{\calP \cap \{y| \ip{y,v} = a\}} \sum_{k\in \Z} e^{-\pi (\frac{k + a}{\|v\|})^2} dx~da \\
 &=& \frac{1}{\|v\|} \int_0^1 \|v\|d(L^*)\cdot \sum_{k\in \Z} e^{-\pi (\frac{k + a}{\|v\|})^2} da \\
  &=&  d(L^*) \int_0^1 \sum_{k\in \Z} e^{-\pi (\frac{k + a}{\|v\|})^2} da
  = d(L^*) \int_{-\infty}^{\infty} e^{-\pi (\frac{a}{\|v\|})^2} da = \|v\| d(L^*).
\end{eqnarray*}

\end{proof}

\begin{claim}\label{joint_statistical_distance}
Let $X_1,\ldots,X_m,Y_1,\ldots,Y_m$ be mutually independent random variables. Then the statistical distance between the
joint distributions satisfies $$\Delta((X_1,\ldots,X_m),(Y_1,\ldots,Y_m)) \le \sum_{i=1}^m \Delta(X_i,Y_i).$$
\end{claim}
\begin{proof}
We consider the case $m=2$. The claim follows for $m>2$ by induction. According to the triangle inequality,
$$ \Delta((X_1,X_2),(Y_1,Y_2)) \le \Delta((X_1,X_2),(X_1,Y_2)) + \Delta((X_1,Y_2),(Y_1,Y_2)).$$
Since $X_1$ is independent of $X_2$ and $Y_2$,
$$\Delta((X_1,X_2),(X_1,Y_2)) = \Delta(X_2,Y_2)$$
and similarly
$$\Delta((X_1,Y_2),(Y_1,Y_2)) = \Delta(X_1,Y_1).$$
\end{proof}

\subsection*{Properties of an LLL reduced basis}

\begin{claim}\label{matrix_inverse}
Let $B=(b_{i,j})_{1\le i,j \le n}$ be an $n\times n$ upper triangular matrix such that for all $i<j\le n$, $|b_{i,j}|
\le |b_{i,i}|$. Then, the entries of $(B^T)^{-1}$ have an absolute value of at most $\frac{1}{\min_i |b_{i,i}|} 2^n$.
\end{claim}
\begin{proof}
First, let $D$ denote the diagonal matrix with values $b_{i,i}$ on the diagonal. Then $B$ can be written as $M D$ where
$M$ is an upper triangular matrix with ones on the diagonal and all other entries have an absolute value of at most
$1$. Then, $(B^T)^{-1}=(D^T M^T)^{-1} = (M^T)^{-1} D^{-1}$. Therefore, it is enough to show that the entries of
$L:=(M^T)^{-1}$ have absolute values of at most $1$. The diagonal of $L$ is all ones and it is lower triangular. The
entry $l_{i,j}$ for $i>j$ can be recursively defined by $-\sum_{j\le k< i} l_{k,j}m_{k,i}$. Therefore,
$$|l_{i,j}| = \abs{\sum_{j\le k< i} l_{k,j}m_{k,i}} \le \sum_{j\le k< i} \abs{l_{k,j}m_{k,i}} \le \sum_{j\le k< i} \abs{l_{k,j}}$$
from which we get the bound $|l_{i,j}|\le 2^{i-j}$ for $i\ge j$.
\end{proof}

\begin{claim}\label{LLL}
Let $(v_1,\ldots,v_n)$ be an $LLL$-reduced basis of a lattice $L$ and let $\sum_{i=1}^n a_i v_i$ be its shortest
vector. Then $|a_i| \le 2^{2n}$ for all $i\in [n]$ and $\lambda(L) \le \|v_1\| \le 2^n \lambda(L)$. Moreover, if
$(v^*_1,\ldots,v^*_n)$ is the dual basis, then $\norm{v^*_i} \le \frac{\sqrt{n}}{\lambda(L)} 2^{2n}$ for all $i\in
[n]$.
\end{claim}
\begin{proof}
Let $(v_1^\dag,\ldots,v_n^\dag)$ denote the Gram-Schmidt orthogonalization of $(v_1,\ldots,v_n)$, i.e., $v_i^\dag$ is
the component of $v_i$ orthogonal to the subspace spanned by $v_1,\ldots,v_{i-1}$. Clearly, $\ip{v^\dag_i,v_j}=0$ for
$i>j$. Recall that in an LLL reduced basis $\norm{v_i^\dag} \le \sqrt{2} \norm{v_{i+1}^\dag}$ and for $i<j$,
$\abs{\ip{v_i^\dag,v_j}} \le \frac{1}{2} \norm{v_i^\dag}^2$. In addition, recall that $\min_i \norm{v_i^\dag}$ is a
lower bound on $\lambda(L)$. Then for any $i\in [n]$, $\norm{v_1^\dag} \le 2^{(i-1)/2} \norm{v_i^\dag}$ and therefore
$\norm{v_1^\dag} \le 2^{(n-1)/2} \lambda(L)$. But since $v_1^\dag = v_1$ we see that $\lambda(L) \le \norm{v_1} \le 2^n
\lambda(L)$. Consider the representation of $(v_1,\ldots,v_n)$ in the orthonormal basis $(v_1^\dag/\norm{v_1^\dag},
\ldots, v_n^\dag/\norm{v_n^\dag})$. It is given by the columns of the matrix $B=(b_{i,j})_{1\le i,j\le n}$ where
$b_{i,j}=\ip{v_j,v_i^\dag}/\norm{v_i^\dag}$. Notice that this matrix is upper triangular and that its diagonal is
$b_{i,i} = \norm{v_i^\dag}$. Also note that by the properties of an LLL reduced basis, $|b_{i,j}| \le
\frac{1}{2}\norm{v_i^\dag}$ for $i<j$. The shortest vector is $\sum_{i=1}^n a_i v_i = \sum_{i=1}^n (\sum_{j=i}^n a_j
b_{i,j}) v_i^\dag / \norm{v_i^\dag}$. Since its length is at most $2^n \norm{ v_i^\dag }$ the absolute value of each of
its coordinates is at most $2^n \norm{ v_i^\dag }$. Hence, $\abs{\sum_{j=i}^n a_j b_{i,j}} \le 2^n \norm{v_i^\dag}$ for
every $i\in [n]$. By taking $i=n$ we get that $|a_n b_{n,n}| \le 2^n \norm{v_n^\dag}$ and hence $|a_n|$ is at most
$2^n$. We continue inductively and show that $|a_k| \le 2^{2n-k}$. Assume that the claim holds for
$a_{k+1},\ldots,a_n$. Then, $|\sum_{j=k+1}^{n} a_j b_{k,j}| \le \frac{1}{2} |\sum_{j=k+1}^n a_j| \|v_k^\dag \| \le
\frac{1}{2} (\sum_{j=k+1}^n 2^{2n-j}) \|v_k^\dag \| \le \frac{1}{2} \cdot 2^{2n-k} \|v_k^\dag \|$. By the triangle
inequality, $|a_k b_{k,k}| \le |\sum_{j=k+1}^{n} a_j b_{k,j}| + |\sum_{j=k}^{n} a_j b_{k,j}| \le (\frac{1}{2} 2^{2n-k}
+ 2^n) \|v_k^\dag \| \le 2^{2n-k}\|v_k^\dag \|$ and the proof of the first part is completed.

The basis of the dual lattice is given by the columns of $(B^T)^{-1}$. Since $\min_i |b_{i,i}| \ge
\frac{\lambda(L)}{2^n}$ and $|b_{i,j}| \le \frac{1}{2} |b_{i,i}|$, Claim~\ref{matrix_inverse} implies that the entries
of $(B^T)^{-1}$ are at most $\frac{1}{\lambda(L)} 2^{2n}$ in absolute value. Therefore, the length of each column
vector is at most $\frac{\sqrt{n}}{\lambda(L)} 2^{2n}$.
\end{proof}

\end{document}